\documentclass[draftclsnofoot, 12pt, onecolumn, hidelinks]{IEEEtran}

\usepackage[utf8]{inputenc}
\usepackage{graphicx}
\usepackage{amsmath,amssymb,mathtools}
\usepackage{bm}
\usepackage{caption}
\usepackage[shortlabels]{enumitem}
\usepackage{cite}
\usepackage{soul}
\usepackage[acronym,shortcuts]{glossaries}
\usepackage{fancyhdr} 
\usepackage{float}
\usepackage{relsize}
\usepackage{outlines}
\usepackage{dsfont}
\usepackage{multirow}
\usepackage{mathtools}
\usepackage{upgreek}
\usepackage{psfrag}
\usepackage{caption}
\usepackage{subcaption}
\usepackage{mathrsfs}
\usepackage{stackengine}
\usepackage{hyperref}
\usepackage{tikz}
\usepackage[normalem]{ulem}
\usepackage{bbm}
\usepackage{pmat}
\usepackage[bottom]{footmisc}
\usepackage{algorithm}
\usepackage[noend]{algpseudocode}
\usepackage{hyperref}

\newcommand{\diag}[1]{{\mathrm{diag}}\!\left({#1}\right)}
\newcommand{\trans}[0]{^{\mathsf{T}}}
\newcommand{\Exp}[0]{\mathbb{E}}
\newcommand{\Var}[0]{\mathrm{Var}}

\newacronym{JCAS}{JCAS}{joint communication and sensing}
\newacronym{RC-JCAS}{RC-JCAS}{radar-centric JCAS}
\newacronym{CC-JCAS}{CC-JCAS}{communication-centric JCAS}
\newacronym{RC-ISAC}{RC-ISAC}{radar-centric ISAC}
\newacronym{CC-ISAC}{CC-ISAC}{communication-centric ISAC}
\newacronym{AP}{AP}{access point}
\newacronym{AD}{AD}{autonomous driving}
\newacronym{DN}{DN}{drone networking}
\newacronym{MIMO}{MIMO}{multiple-input multiple-output}
\newacronym{DoF}{DoF}{degree of freedom}
\newacronym{CPU}{CPU}{central processing unit}
\newacronym{ROI}{ROI}{region of interest}
\newacronym{LOS}{LOS}{line-of-sight}
\newacronym{VOER}{VOER}{voxel-occupancy-error-rate}
\newacronym{VCER}{VCER}{voxel-coefficient-error-rate}
\newacronym{VER}{VER}{voxel-error-rate}
\newacronym{NLOS}{NLOS}{non-line-of-sight}
\newacronym{AWGN}{AWGN}{additive white Gaussian noise}
\newacronym{3D}{3D}{three-dimensional}
\newacronym{GaBP}{GaBP}{Gaussian belief propagation}
\newacronym{BiGaBP}{BiGaBP}{bilinear Gaussian belief propagation}
\newacronym{BiGAMP}{BiGAMP}{bilinear generalized approximate message passing}
\newacronym{IC}{IC}{interference cancellation}
\newacronym{radar}{radar}{radio detection and ranging}
\newacronym{SE}{SE}{spectral efficiency}
\newacronym{RCS}{RCS}{radar cross-section}
\newacronym{MP}{MP}{message passing}
\newacronym{MSE}{MSE}{mean-squared-error}
\newacronym{PDF}{PDF}{probability distribution function}
\newacronym{AoA}{AoA}{angle-of-arrival}
\newacronym{OTFS}{OTFS}{orthogonal time frequency space}
\newacronym{OFDM}{OFDM}{orthogonal frequency-division multiplex}
\newacronym{SER}{SER}{symbol-error-rate}
\newacronym{CSI}{CSI}{channel state information}
\newacronym{CLT}{CLT}{central limit theorem}
\newacronym{SGA}{SGA}{scalar Gaussian approximation}
\newacronym{AMP}{AMP}{approximate message passing}
\newacronym{IoT}{IoT}{Internet-of-Things}
\newacronym{ISAC}{ISAC}{integrated sensing and communications}
\newacronym{B5G}{B5G}{beyond fifth-generation}
\newacronym{6G}{6G}{sixth-generation}
\newacronym{BER}{BER}{bit-error-rate}
\newacronym{URLLC}{URLLC}{ultra-reliable low-latency communications}
\newacronym{mMTC}{mMTC}{massive machine-type communications}
\newacronym{mMIMO}{mMIMO}{massive multiple-input multiple-output}
\newacronym{mmWave}{mmWave}{millimeter-wave}
\newacronym{THz}{THz}{Terahertz}
\newacronym{RIS}{RIS}{reconfigurable intelligent surface}
\newacronym{IRS}{IRS}{intelligent reflective surface}
\newacronym{FD}{FD}{full duplex}
\newacronym{CR}{CR}{cognitive radio}
\newacronym{RCC}{RCC}{radar and communication coexistence}
\newacronym{DFRC}{DFRC}{dual-function radar communications}
\newacronym{IM}{IM}{index modulation}
\newacronym{GAMP}{GAMP}{generalized approximate message passing}
\newacronym{SCMA}{SCMA}{sparse code multiple access}
\newacronym{CS}{CS}{compressive sensing}
\newacronym{FMCW}{FMCW}{frequency modulated continuous waveforms}
\newacronym{i.i.d.}{i.i.d.}{independent and identically distributed}
\newacronym{CAESAR}{CAESAR}{carrier agile phased array radar}
\newacronym{SotA}{SotA}{state-of-the-art}
\newacronym{AL-ISAC}{AL-ISAC}{Alternating Linear ISAC}
\newacronym{Bi-ISAC}{Bi-ISAC}{bilinear ISAC}
\newacronym{SNR}{SNR}{signal-to-noise ratio}
\newacronym{MQAM}{$M$-QAM}{$M$-ary quadrature amplitude modulation}
\newacronym{flop}{flop}{floating-point operation}
\newacronym{FP}{FP}{false-positive}
\newacronym{FN}{FN}{false-negative}

\newglossaryentry{UE}
{
name={UE},
description={user equipment},
first={\glsentrydesc{UE} (\glsentrytext{UE})},
plural={UEs},
firstplural={user equipment (\glsentryplural{UE})}
}

\title{Integrated Sensing and Communications \\[-0.5ex] for 3D Object Imaging via Bilinear Inference}

\author{
Hyeon Seok Rou, \IEEEmembership{Graduate Student Member, IEEE}, \\[-0.5ex] 
Giuseppe Thadeu Freitas de Abreu, \IEEEmembership{Senior Member, IEEE},  \\[-0.5ex] 
David Gonz{\'a}lez G., \IEEEmembership{Senior Member, IEEE}, and Osvaldo Gonsa

\thanks{\noindent H.~S.~Rou and G.~T.~F.~Abreu are with the School of Computer Science and Engineering, Constructor University (\emph{previously Jacobs University Bremen}), Campus Ring 1, 28759, Bremen, Germany (e-mails: [hrou, gabreu]@constructor.university).}
\thanks{D.~Gonz{\'a}lez~G. and O.~Gonsa are with the Wireless Communications Technologies Group, Continental AG, Wilhelm-Fay Str. 30, 65936, Frankfurt am Main, Germany (e-mails: [david.gonzalez.gonzalez,  osvaldo.gonsa]@continental-corporation.com).}
\thanks{A part of this article has been presented at the {\color{black}2022} 56th Asilomar Conference on Signals, Systems, and Computers, \cite{Rou_Asilomar22JCAS}.}
}

\begin{document}

\maketitle

\vspace{1ex}
\vspace{-9.5ex}
\begin{abstract}
\vspace{-1ex}
%
We consider an uplink \ac{ISAC} scenario where the detection of data symbols from multiple \acp{UE} occurs simultaneously with a \ac{3D} estimation of the environment, extracted from the scattering features present in the \ac{CSI} and utilizing the same physical layer communications air interface, as opposed to radar technologies.
By exploiting a discrete (voxelated) representation of the environment, two novel \ac{ISAC} schemes are derived with purpose-built \ac{MP} rules for the joint estimation of data symbols and status (filled/empty) of the discretized environment.
The first relies on a modular feedback structure in which the data symbols and the environment are estimated alternately, whereas the second leverages a bilinear inference framework to estimate both variables concurrently.
Both contributed methods are shown via simulations to outperform the \ac{SotA} in accurately recovering the transmitted data as well as the \ac{3D} image of the environment. 
An analysis of the computational complexities of the proposed methods reveals distinct advantages of each scheme, namely, that the bilinear solution exhibits a {\color{black}superior robustness to short pilots and channel blockages,} while the {\color{black}alternating} solution offers lower complexity with large number of \acp{UE} {\color{black}and superior performance in ideal conditions.}


\end{abstract}
\glsresetall

\vspace{-2ex}
\section{Introduction}
\label{sec:intro}
\vspace{-0.5ex}

{\color{black}
\Ac{B5G} and \ac{6G} wireless communication systems are expected to raise the performance standards} in terms of data throughput, reliability, latency, spectral efficiency, energy efficiency, and user capacity \cite{Saad_Network20, Zhang_VTM19, Rappaport_TAP17}, which can be achieved by novel enabling technologies such as high-frequency communications in the \ac{mmWave} \cite{ Wang_CST18, Xiao_JSAC17, Niu_Networks15} and \ac{THz} \cite{Song_TTHz22, Elayan_CommNet18, Rappaport_Access19} bands, \ac{mMIMO} techniques \cite{Zhang_Access19, Carvalho_WC20, Wang_JSTSP19}, \acp{RIS} \cite{Basar_Access19, Wu_TWC19, Huang_TWC19}, and more.

Within that context, {\color{black}a new research field called} \ac{ISAC} \cite{Liu_JSAC22, Liu_CST22, Xiao_WCSP21}, also known as \ac{JCAS} \cite{Thoma_EuCAP21, Wild_Access21, Zhang_JSTSP21, Zhang_CST22, Liu_TC20}, {\color{black}has recently gained significant attention as a promising technology to fulfill such requirements and enable new applications for \ac{B5G} and \ac{6G} systems.
In particular, \ac{ISAC} technology seeks to enhance \Ac{B5G} and \ac{6G} systems by enabling both communication and environment sensing functionalities under \ul{the same} wireless interface, thus realizing  the concepts of ambient-sensing and environment-aware radio} \cite{ZengWComm2021}, which {\color{black}are} crucial to emerging applications such as \ac{AD} \cite{Al-DulaimiComSM2020} and \ac{DN} \cite{MishraComSM2022}{\color{black}, besides offering new means to optimize system performance.}
For instance, in \ac{B5G} and \ac{6G} systems operating at high-frequency channels, which are sensitive to path-dependent scattering \cite{Xing_GLOBECOM19, MyListOfPapers:KanaunBook2020}, environment parameters of interest include not only the ``usual'' \ac{CSI}, but also the positions of users and obstacles that may lead to path blockages.
In such systems, \ac{ISAC} is an alternative to image-based path-blockage prediction approaches, crucial to mitigate the deleterious effects of blockages \cite{Charan2021, Yu2022, Wu2022}.

{\color{black}
The prominent challenge of \ac{ISAC} arises from the fact that two independently-developed wireless technologies, namely, wireless communications and radar systems \cite{MyListOfPapers:RadarSigProc2014, Long_CIS19}, are both fundamentally based on distinct air-interfaces, such that a concurrent deployment of existing waveforms is bound to suffer from performance degradation of both functions due to interference.
Aiming to frontally address this issue, the earliest family of \ac{ISAC} approaches known as the \ac{RCC} \cite{Zheng_SPM19,Liu_TSP18}, consequently focused on} minimizing the interference and maximizing the cooperation between the independently-operated {\color{black}communications and radar} subsystems sharing the same frequency spectrum.
{\color{black} While the \ac{RCC} strategy addresses succeeds in managing interference and harmoniously allocating radio resources to operate both subsystems, the approach achieves relatively low spectral efficiencies and does not alleviate hardware costs, whose components remain separate for both subsystems \cite{Zheng_SPM19,Zhang_CST22, Liu_TSP18}.}

{\color{black} In light of the fundamental drawbacks of \ac{RCC}, techniques integrating the communication and radar functions into a single wireless interface have been recently proposed to truly realize the \ac{ISAC} goal of jointly offering communication and sensing capabilities in a single system.
Literature \cite{Zhang_JSTSP21,Zhang_CST22,Liu_TC20} classifies such techniques} into three types:
\textit{a)} {\color{black} {\bf \Ac{RC-ISAC}} schemes, which realizes an additional communication functionality over typical radar waveforms,} for example by utilizing \ac{IM} to encode information into \ac{MIMO} radar signals employing,  \textit{e.g.}, the \ac{CAESAR} waveform \cite{Huang_TSP20}, or the \ac{FMCW} \cite{Ma_JSTSP21};
\textit{b)} {\color{black} {\bf \Ac{CC-ISAC}} schemes, which realizes an additional environment sensing functionality over standard communication waveforms, typically by extracting the radar parameters such as Doppler-shift and delay from waveforms designed fundamentally for communications functions, such as} the IEEE~802.11ad waveform{\color{black} \cite{Kumari_TVT18}, the} \ac{OFDM} waveform \cite{Liu_RADAR16}, or the \ac{OTFS} waveform \cite{Raviteja_Radarconf19};
and \textit{c)} {\bf\Ac{DFRC}} schemes, {\color{black}which while not having exclusive boundaries with aforementioned \ac{RC-ISAC} and \ac{CC-ISAC} categories, are based on waveforms that can be adaptively or jointly optimized between the two functionalities \cite{Yuan_TVT21}, \textit{e.g.,} via waveforms designed based on mutual information \cite{Liu_CL17} or the multi-beam approach in the mmWave bands \cite{Luo_TVT20}.}
But still, all these approaches are related by the fact that the target sensing is enabled by the {\color{black}fundamental radar relationship between measurable physical quantities and information on the target \cite{MyListOfPapers:RadarSigProc2014,Long_CIS19}, that is, target range can be extracted from the delay in the signal, velocity from the Doppler frequency, and bearing from the \ac{AoA}.} {

Concomitant with the aforementioned methods, contributions have also been recently made to realize environment sensing capabilities {\color{black}not via target detection with radar, but rather by new standards of environment sensing information such as ambient human activity \cite{Tan_CM18, Yousefi_CM17}, and \ac{3D} environment image \cite{Tong_JSTSP21,Tao_WCSP20, Zhang_ICC2022}.
In particular, the latter family of works exploit the voxelated occupancy grid \cite{TropkinaAWPL2022, MoralesTITS2017,Pirker_ICCVW11, Collins_MCCA07} to discretize the environment into \ac{3D} cubic units of space representing its state (solid or void).
Such methods exhibit a unique advantage in that the extracted information not only describes the location of objects, but also their \ac{3D} shape and orientation, in any desired resolution according to the voxelated model, enabling useful applications such as \ac{3D} environment mapping and ray-tracing propagation modeling \cite{HeCST2019,Choi_ISAP21}.

However, wireless voxelated imaging technology is still a very new notion in context of \ac{ISAC}, due to the inherently convoluted channel paths arising from the discretization of the environment scatterers, and the fundamental challenge that the unknown information symbols must be \ul{simultaneously} recovered on top of the very large number of environment voxels. }

{\color{black}In light of the above challenges, we offer in this article the following contributions:}
\begin{itemize}
\item An extension of the discrete voxelated environment model utilized in \cite{Tong_JSTSP21,Tao_WCSP20, Zhang_ICC2022} is introduced, in which an empirical stochastic-geometric approach is incorporated to capture the viability of \ac{NLOS} paths, in addition to an extension where the occupancy coefficients are not limited to 0 or 1, but instead can take on non-binary complex values, enabling reflection losses and phase shifts of reflected waves to be modeled\footnote{We clarify that although the message-passing rules derived in this article are for this extended paradigm, such that the proposed algorithms are fully generalized, binary-valued occupancy coefficients are considered in Section \ref{sec:performance} for the purpose of evaluation performance, in order to enable direct comparison with \ac{SotA} methods.}.
\item A novel, scalable \ac{CC-ISAC} scheme is proposed, in which the \ac{3D} voxelated environment image and the transmit symbols are estimated \ul{alternately} via two distinct linear modules; 
\item A novel, high-performance \ac{CC-ISAC} scheme is proposed, in which the the  \ac{3D} voxelated image of the environment and the transmit data symbols are \ul{concurrently} estimated via a single \ac{MP} module which leverages a bilinear inference framework;
\item {\color{black}Insights on the advantages of the two proposed \ac{CC-ISAC} algorithms are provided via performance assessment and comparison against the \acf{SotA}, which highlights the \ul{robustness} of the bilinear method against short pilot lengths and path blockages, and the \ul{accuracy} of the alternating solution in systems with many \acp{UE}.}
\end{itemize}
}

\emph{Notation}: Scalar values are denoted by slanted lowercase letters, as in $x$, while complex vectors and matrices are denoted by boldface lowercase and uppercase letters, as in $\mathbf{x}$ and $\mathbf{X}$, respectively. 
The transposition, complex conjugation, diagonalization, absolute value, and $\ell$-th norm operators are denoted by $(\cdot)\trans$, $(\cdot)^*$, $\diag{\cdot}$, $|\cdot|$, and $||\cdot||_{\ell}$ respectively{, while} $\mathbb{E}_{\mathsf{x}}(x)$ and $\mathrm{Var}_{\mathsf{x}}(x)$ respectively denote the expectation and variance of $x$ with respect to {its} distribution $\mathbb{P}_{\mathsf{x}}(x)$.
The sets of real and complex numbers are denoted by $\mathbb{R}$ and $\mathbb{C}${,} respectively, and $\mathcal{N}(\mu, \nu)$ and $\mathcal{CN}(\mu, \nu)$ denote the real and complex Normal distributions with mean $\mu$ and variance $\nu$.


\vspace{1ex}
\section{System Model}
\label{sec:system_model}

The system model considered throughout the article, consists of three parts: \textit{a)} the environment model, where a voxelated occupancy grid is used to discretely approximate the true environment and its scattering properties, \textit{b)} the channel model, which defines the unique channel paths arising from the voxelated environment model, and \textit{c)} the signal model, where the uplink communication scheme between the \acp{UE} and the \acp{AP} is described.

\vspace{-1ex}
\subsection{Environment Voxelation Model}
\label{sec:voxelation_model}

\begin{figure}[H]
\vspace{-2ex}
\centering
\begin{subfigure}[t]{0.3\textwidth}
\centering
\includegraphics[width=\textwidth]{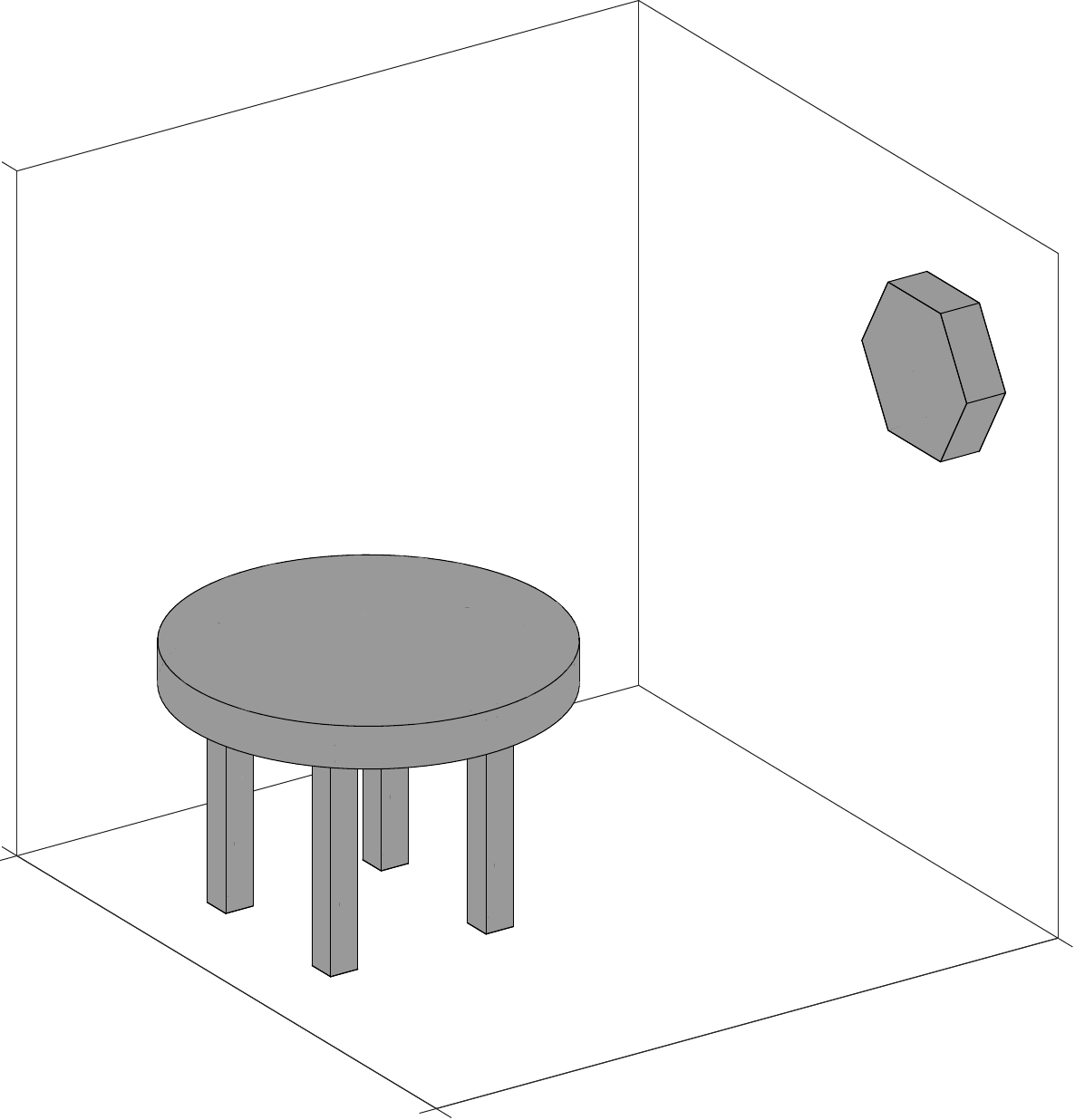}
\caption{True environment ROI.}
\label{fig:env_model_true}
\end{subfigure}
\hfill
\begin{subfigure}[t]{0.3\textwidth}
\centering
\includegraphics[width=\textwidth]{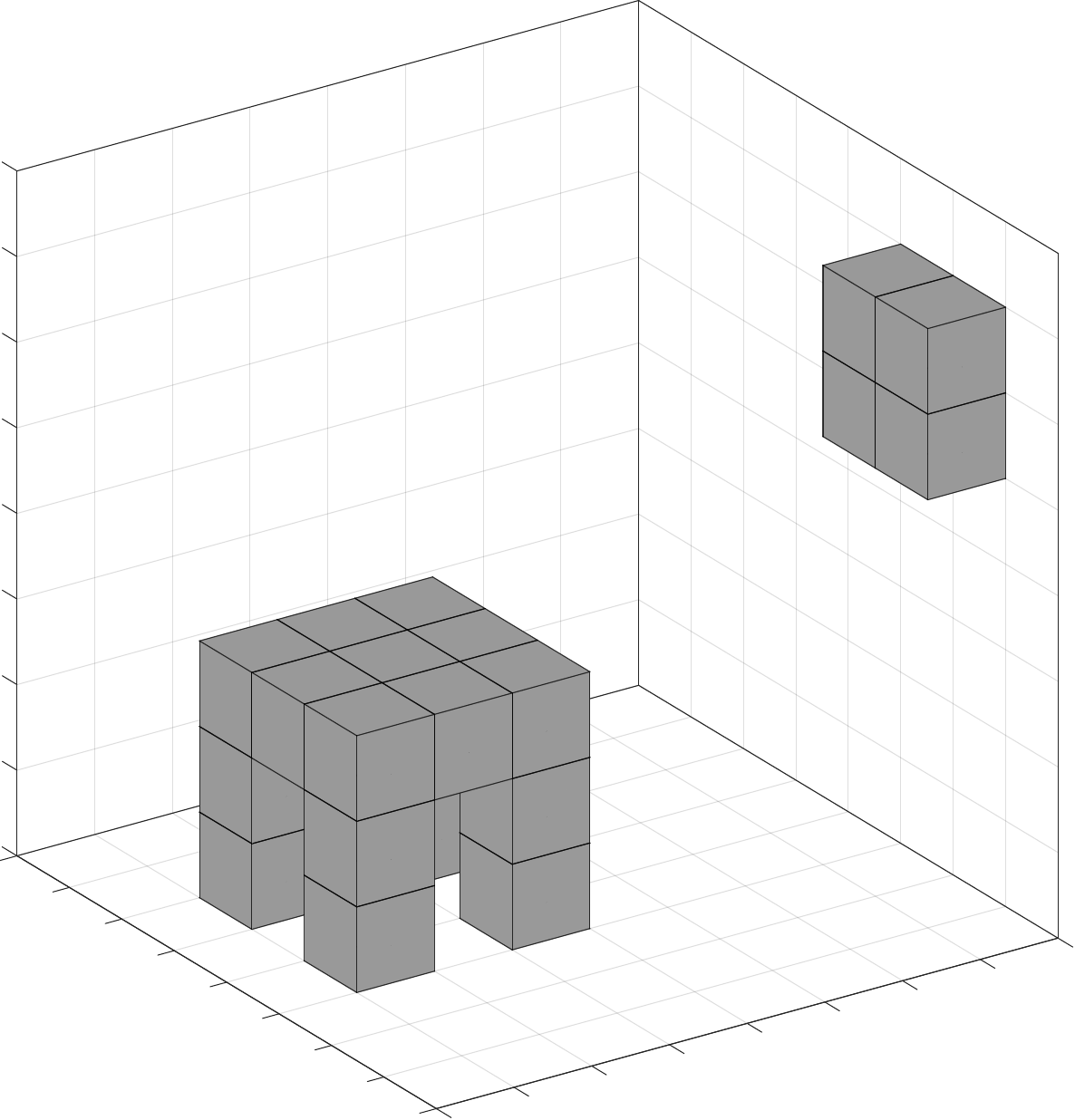}
\caption{Low resolution ROI model.}
\label{fig:env_model1}
\end{subfigure}
\hfill
\begin{subfigure}[t]{0.3\textwidth}
\centering
\includegraphics[width=\textwidth]{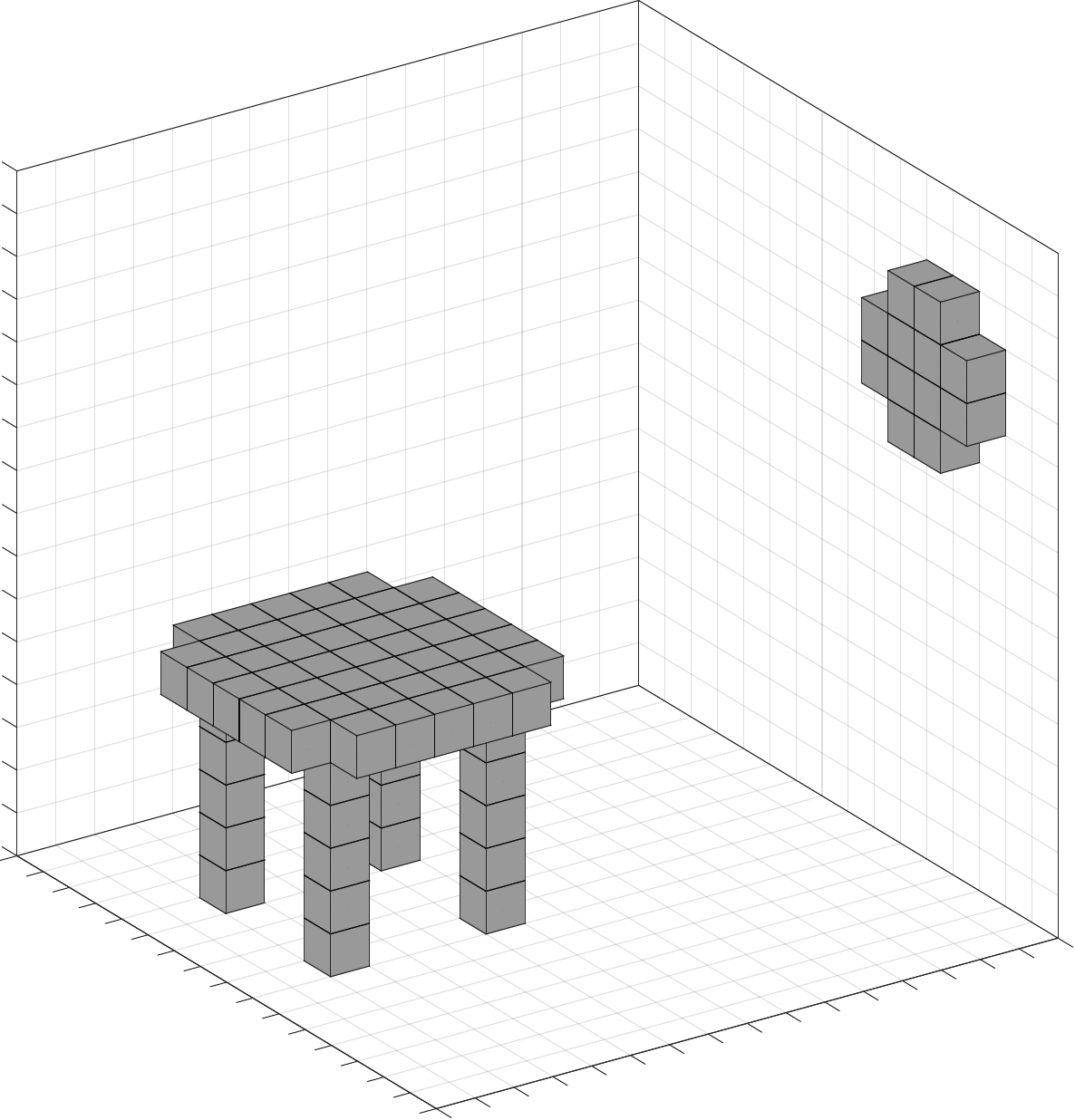}
\caption{High resolution ROI model.}
\label{fig:env_model2}
\end{subfigure}
\vspace{-2ex}
\caption{Illustration of a voxelated grid map-based model of the environment of a given \ac{ROI}.}
\label{fig:voxel_model}
\vspace{-2ex}
\end{figure}

The \ac{3D} image of an environment {can} be represented via a {number} of techniques, including the well-known point-cloud {and} the ray-tracing method{s}, which are {often} utilized in robotics, machine vision and computer graphics \cite{HeCST2019, TuICRA2019}.
Another well-known method, however, is the voxelated occupancy grid \cite{TropkinaAWPL2022, MoralesTITS2017,Pirker_ICCVW11, Collins_MCCA07}, where the \ac{ROI} is {represented} as a cuboidal space of dimensions $L_{\mathrm{x}} \times L_{\mathrm{y}} \times L_{\mathrm{z}}$, each denoting the lengths of the $\mathrm{x}, \mathrm{y}, \mathrm{z}$-axes in meters, respectively, as depicted in Figure \ref{fig:voxel_model}.
In this model, the \ac{ROI} is subdivided into a regular grid consisting of $N_V \triangleq N_{\mathrm{x}} \cdot N_{\mathrm{y}} \cdot N_{\mathrm{z}}$ voxels, where $N_{\mathrm{x}} \triangleq \frac{L_{\mathrm{x}}}{L_V}$, $N_{\mathrm{y}} \triangleq \frac{L_{\mathrm{y}}}{L_V}$, and $N_{\mathrm{z}} \triangleq \frac{L_{\mathrm{z}}}{L_V}$ denote the number of partitions along the $\mathrm{x}, \mathrm{y}$ and $\mathrm{z}$-axes, respectively, and $L_V$ is the edge length of a voxel in meters, which therefore corresponds to the image resolution.

The environment is then represented by a \ac{3D} tensor of dimensions ($N_{\mathrm{x}} \times N_{\mathrm{y}} \times N_{\mathrm{z}}$), whose elements indicate the occupancy of the voxels, and thus whether that portion of the space is empty or filled with a given material.
One may therefore consider, in general, each $k$-th voxel to be represented by an occupancy (a.k.a. scattering) coefficient $v_k$, with $k \in \{1, \cdots, N_V\}$, where $v_k = 0$ indicates that the $k$-th voxel is empty{, while an occupied voxel is indicated by a complex number} \textit{i.e.,} $v_k \triangleq \beta_k \cdot e^{-j\omega_k} \in \mathbb{C}$.
{In such a model, the constants $\beta_k$ and $\omega_k$, which dependent not only on the material itself, but also on the frequency and the angle of incidence of propagating signals \cite{MyListOfPapers:KanaunBook2020},} capture the effect {of the material occupying a given voxel onto} the electromagnetic wave reflected {or refracted by it}.
For the sake of reducing the complexity of the \ac{ISAC} algorithm to be later introduced, however, we will in this article consider a simplified model whereby the occupancy coefficients $v_k$ take on discrete real values in the interval $\in [0,1]$.

Since phase rotations due to reflections are captured by channel estimation, this simplification is equivalent to the assumptions that the \ac{ISAC} waveform is narrowband, so that frequency-dependence can be ignored \cite{MyListOfPapers:KanaunBook2020}.
The incorporation of the geometry of the interaction between propagating waves and occupied voxels will be described in Subsection \ref{sec:environment_model}.

\vspace{-1.5ex}
\subsection{Channel Model}
\label{sec:channel_model}
\vspace{-0.5ex}

\begin{figure}[H]
\centering
\includegraphics[width=0.5\textwidth]{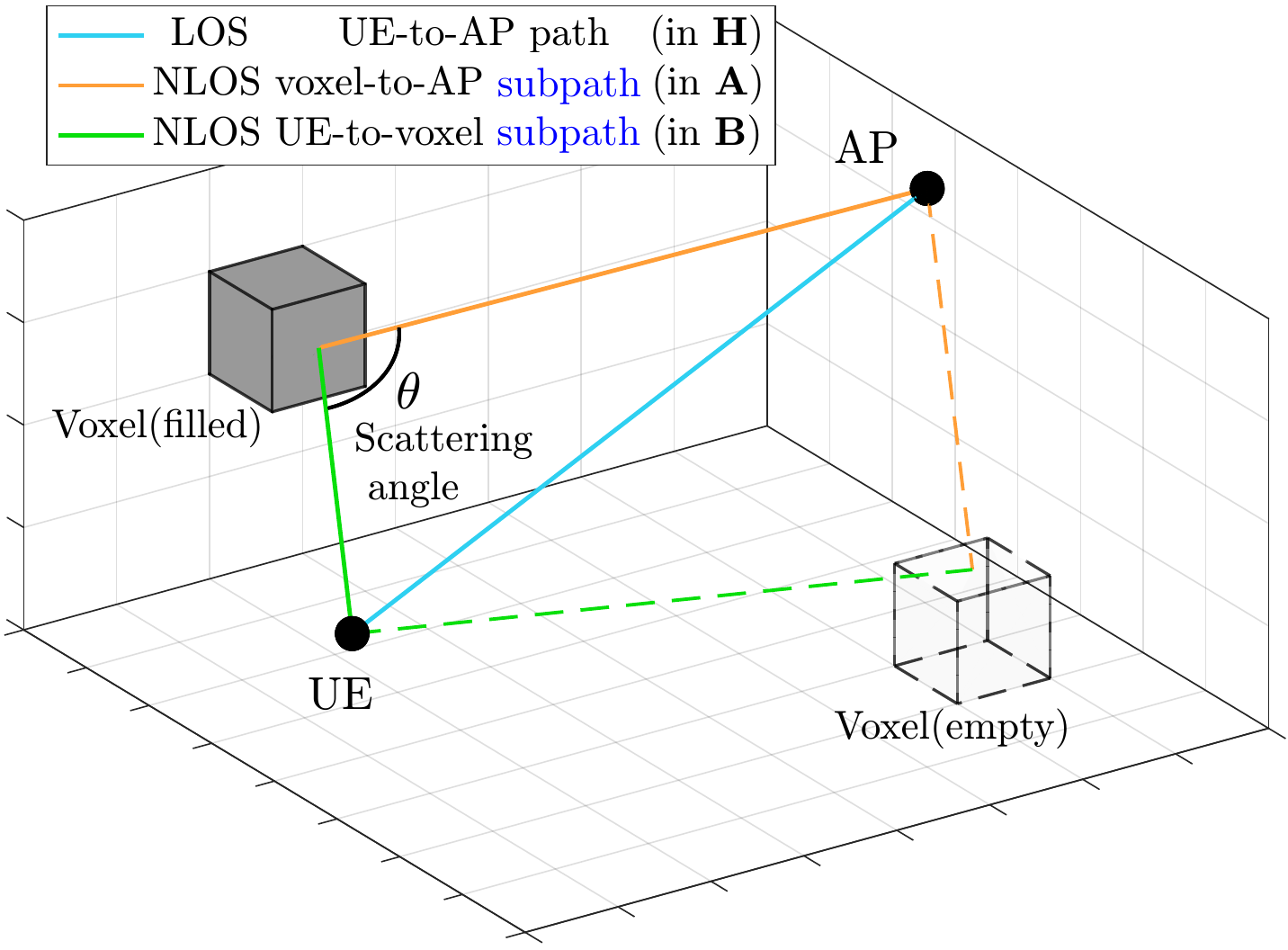}
\vspace{-1ex}
\caption{Illustration of the \ac{LOS} and \ac{NLOS} channel components {\color{black} and their subpaths}.}
\label{fig:path_types}
\vspace{-4.5ex}
\end{figure}

Consider {a scenario in which the \ac{ROI} contains} $N_U$ single-antenna \acp{UE}, and $N_A$ multi-antenna \acp{AP} equipped with $N_R$ receive antennas each.
As illustrated in Fig. \ref{fig:path_types}, the effective channels between the \acp{UE} and \acp{AP} contain two distinct types of components, namely, \ac{LOS} components which are direct paths between the \acp{UE} and \acp{AP}, and \ac{NLOS} components which encompass paths reflected at occupied voxels corresponding to parts of the environment, as described in Subsection \ref{sec:voxelation_model}.
Assuming that the operating frequency band is sufficiently high that the power of paths reflected more than once is negligible \cite{Xing_GLOBECOM19, mmWave_and_THz_Channels_Access2019}, \ac{NLOS} components may be decomposed into two {\color{black}\textit{subpaths}}, the \ac{UE}-to-voxel subpath and the voxel-to-\ac{AP} subpath, which together with the voxel scattering coefficient comprises the aggregate \ac{NLOS} channel.

In light of the above, the effective channel between the $N_U$ {single-antenna} \acp{UE} and {the ensemble of} $N_A N_R$ receive antennas {of all} \acp{AP} {can be compactly described} by
\begin{equation}
\mathbf{G}  =  \hspace{-0.5ex}\overbrace{\mathbf{H}}^{\text{UE-to-AP}} \hspace{-0.5ex}+ \hspace{-1ex}\overbrace{\mathbf{A}}^{\text{voxel-to-AP}} \hspace{-2.5ex}\cdot \hspace{0.5ex} \mathrm{diag}({\mathbf{v}})\cdot\hspace{-2.5ex} \overbrace{\mathbf{B}}^{\text{UE-to-voxel}} \hspace{-1ex} \in \mathbb{C}^{N_A N_R \times N_U},
\label{eq:effective_channel}
\vspace{-1ex}
\end{equation}
where $\mathbf{G} \in \mathbb{C}^{N_A N_R \times N_U}$ is the effective channel matrix, $\mathbf{H} \in \mathbb{C}^{N_A N_R \times N_U}$, $\mathbf{A} \in \mathbb{C}^{N_A N_R \times N_V}$, and  $\mathbf{B} \in \mathbb{C}^{N_V \times N_U}$ are the  channel matrices of the \ac{UE}-to-\ac{AP} \ac{LOS} path, voxel-to-\ac{AP} \ac{NLOS} subpath, and \ac{UE}-to-voxel \ac{NLOS} subpath, respectively, whose elements are assumed to follow zero-mean complex Normal distribution{s} with variances $\sigma_H^2, \sigma_A^2$, and $\sigma_B^2$, respectively; {while} $\mathbf{v} \in \mathbb{C}^{N_V \times 1}$ is the vector containing all scattering coefficients of the voxelated grid.

{\color{black} Although not further exploited in this article, we emphasize that} the channel model in equation \eqref{eq:effective_channel} can {\color{black}be straightforwardly} extended to a multi-carrier scenario {\color{black} by simply introducing frequency-selectivity such that the environment variables are functions of the carrier frequency, \textit{i.e.,}}
\vspace{-1ex}
\begin{equation}
\mathbf{G}(f) = \mathbf{H}(f) + \mathbf{A}(f) \cdot \diag{\mathbf{v}(f)} \cdot \mathbf{B}(f) ~\forall f \in \mathcal{F},
\vspace{-1ex}
\label{eq:effective_channelMultiCarrier}
\end{equation}
with $f$ denoting a specific frequency in the set $\mathcal{F}$ of all subcarrier frequencies. 

{\color{black} We leave details for follow-up work, but it shall become evident} that under such an extended model, the sensing component of the \ac{ISAC} algorithm to be introduced in Section \ref{sec:proposed_solution} can also be extended for even better performance, for instance by exploiting knowledge of channel correlation across carriers \cite{MyListOfPapers:KangThesis2005}, and to incorporate additional features, such as the estimation of material types based on the frequency-dependence observed on estimated scattering coefficients \cite{MyListOfPapers:KanaunBook2020}.
{\color{black} Since the above also requires incorporating message-passing rules that exploit cross-carrier correlation to the algorithms,} such extension will be pursued in a future work, and {\color{black} only the frequency-independent ($i.e.$, single-carrier) model of equation \eqref{eq:effective_channel} will be assumed in this article.}

\vspace{-1.5ex}
\subsection{Stochastic Geometric Environment Model}
\label{sec:environment_model}
\vspace{-0.8ex}

Notice that the channel model summarized by equation \eqref{eq:effective_channel} implies that all paths between \acp{UE}, \acp{AP}, and voxels are available.
Although such an assumption is common in related literature (see $e.g.$ \cite{Tong_JSTSP21,Tao_WCSP20, Zhang_ICC2022, TropkinaAWPL2022, MoralesTITS2017, Pirker_ICCVW11, Collins_MCCA07}),
in practice, many subpaths may not be available due to either physical phenomena ($e.g.$ blockage by air-borne particles, absorption, or path loss) or the finite resolution of the voxelated model itself. 
In order to capture such realistic behavior, the work in \cite{Zhang_ICC2022} considers the occlusion effect of waves reaching the voxels, where the \ac{UE}-to-voxel subpaths are assumed to be unavailable if a voxel is present nearby the path.
This perturbation effect was approximated by applying a \ac{3D} Gaussian kernel convolution to the channel matrix, but it was acknowledged \cite{Zhang_ICC2022} that this approach is a highly simplified model of the true physical phenomena.
Building on the latter, we therefore seek to contribute to improving the voxelated grid model by considering the \ul{statistical feasibility} of paths.

\begin{figure}[t]
\centering
\begin{subfigure}{0.45\textwidth}
\centering
\includegraphics[width=\textwidth]{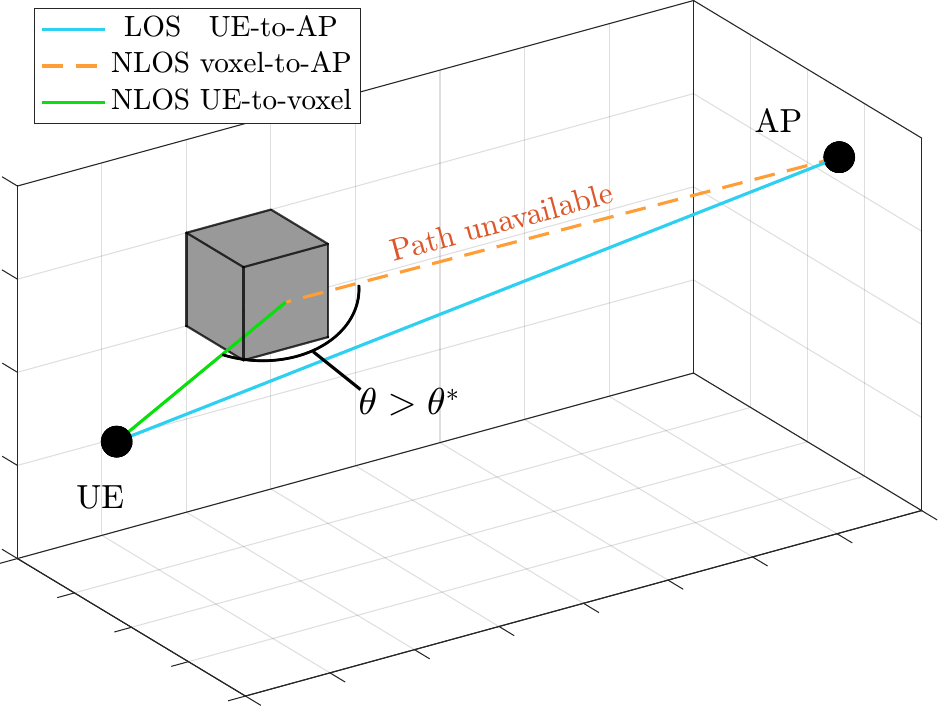}
\caption{Path unavailability due to obtuse scattering.}
\label{fig:infeasible_angle1}
\end{subfigure}
\hfill
\begin{subfigure}{0.45\textwidth}
\centering
\includegraphics[width=\textwidth]{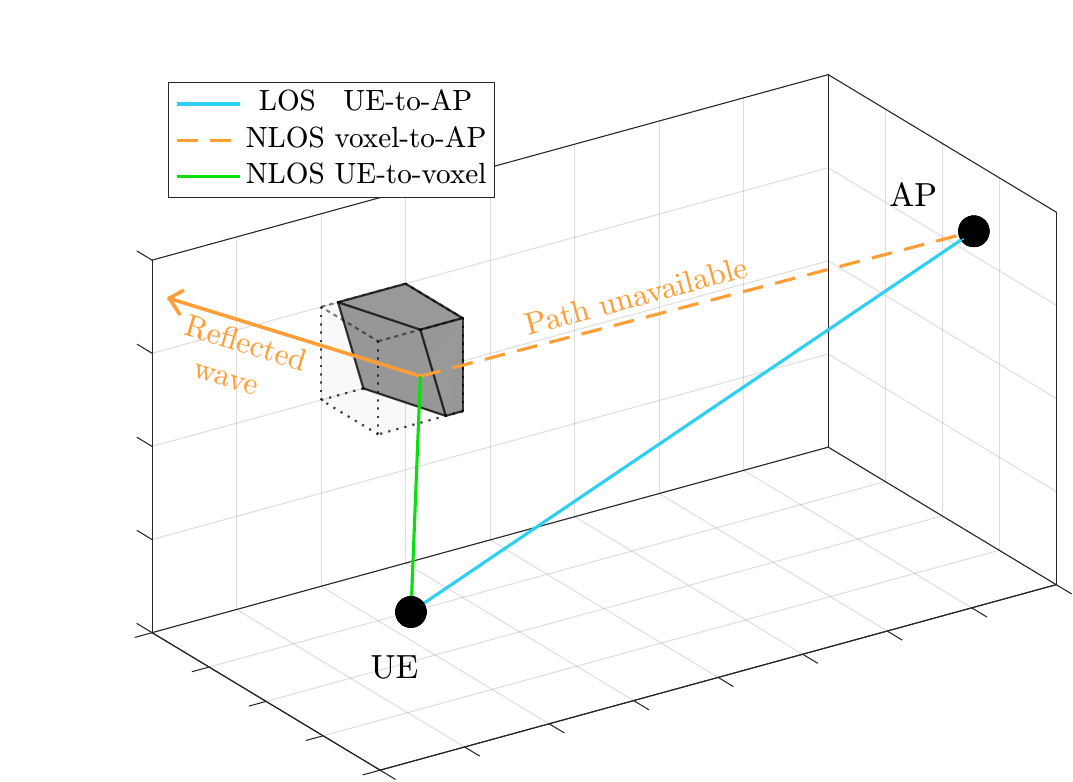}
\caption{Path unavailability due to skewed surface.}
\label{fig:infeasible_angle2}
\end{subfigure}
\vspace{-1.5ex}
\caption{Illustration of physical phenomena leading to the unavailability of propagation paths.}
\label{fig:infeasible_angle}
\vspace{-4.5ex}
\end{figure}

To this end, we refer to the physical phenomena occurring at the reflection of propagating waves, in particular, the fact that for any given frequency: a) a critical angle $\theta^*$ exists such that, as illustrated in Fig. \ref{fig:infeasible_angle1}, if the incidence angle $\theta > \theta^*$, the wave is absorbed rather than reflected, and consequently the corresponding voxel-to-\ac{AP} \ac{NLOS} subpath is not available  \cite{MyListOfPapers:KanaunBook2020}; and b) the curvature of the surface exposed to the impinging wave may be such that no signal is reflected towards an \ac{AP}{\color{black}\footnote{\color{black} Although the phenomenon in Fig. \ref{fig:infeasible_angle2} would reduce to the phenomenon in Fig. \ref{fig:infeasible_angle1} for infinitely small voxels, such extreme resolution leads to prohibitive complexity of the algorithms, so that modeling both phenomena distinctly is preferred in practice.}} \cite{TropkinaAWPL2022}, as illustrated in Fig. \ref{fig:infeasible_angle2}.
But since the complexity of modeling such phenomena at each voxel is far too complex to carry out, especially if the resolution of the voxelated \ac{ROI} is large, we instead employ a statistical approach whereby the angle between the impinging and reflected waves at each voxel, hereafter referred to as the \emph{scattering angle}, and consequently, the availability of each voxel-to-\ac{AP} \ac{NLOS} subpath, are considered.

In light of the above, the following stochastic-geometric model is proposed to integrate the aforementioned phenomena into the channel matrices of the voxelated grid environment model.
First, the positions of the \acp{UE} and the \acp{AP} are discretized into the \ac{3D} grid of the voxelated \ac{ROI}, such that their positions may be described by voxel coordinates\footnote{It is also assumed that the multiple antennas of the \acp{AP} are placed within a single voxel, such that their \ac{AoA} are assumed to be identical, albeit each with a different channel path coefficient.}
Denoting the \ac{3D} coordinates of {an} \ac{UE}, {an} \ac{AP}, and an environment voxel{, respectively by} $\mathbf{c}_{U} = [x_{U}, y_{U}, z_{U}]\trans \in \mathbb{R}^{3}$, $\mathbf{c}_{A} = [x_{A}, y_{A}, z_{A}]\trans \in \mathbb{R}^{3}$, and $\mathbf{c}_{V} = [x_{V}, y_{V}, z_{V}]\trans \in \mathbb{R}^{3}$, the scattering angle $\theta$ of the {\ac{NLOS}} path {reflected} at the voxel {is given by} \vspace{-0.5ex}
\begin{equation}
\theta = \mathrm{arccos}\!\left( \dfrac{(\mathbf{c}_U - \mathbf{c}_V)\trans (\mathbf{c}_A - \mathbf{c}_V)}{||\mathbf{c}_U - \mathbf{c}_V||\!\cdot\!||\mathbf{c}_A - \mathbf{c}_V||} \right) \in [0, \pi],
\label{eq:scatter_angle}
\end{equation}
where $\mathrm{arccos}(\cdot)$ denotes the inverse cosine trigonometric function. 

\begin{figure}[H]
\centering
\includegraphics[width=1\textwidth]{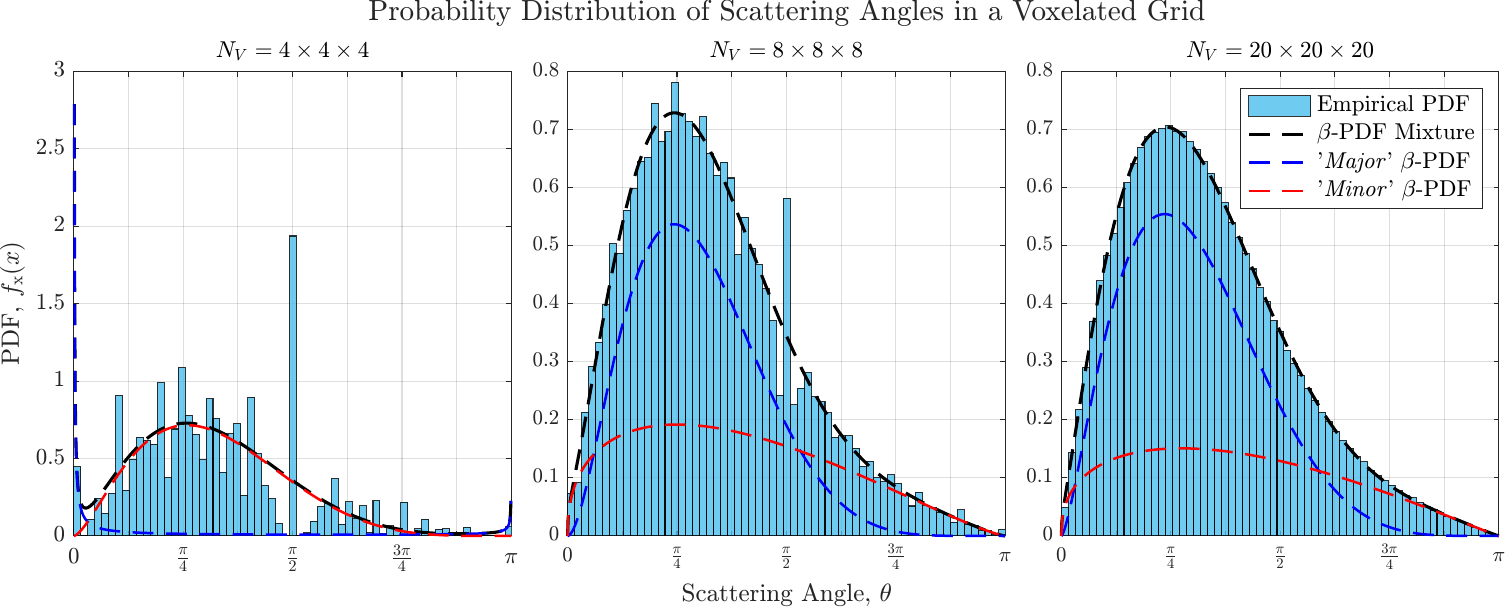}
\vspace{-5.5ex}
\caption{Empirical beta mixture modeling of scattering angle distributions.}
\vspace{-1ex}
\label{fig:angle_dist_PDF}
\vspace{-2ex}
\end{figure}

The empirical\footnote{In principle, the analytical distribution of scattering angles in eq. \eqref{eq:scatter_angle} can be derived, either by studying the $\binom{N_V}{3}$ constituent angles within the highly subdivided geometry of the grid, or via a stochastic geometry-based grid analysis \cite{MillerStochastic07}.
To the best of the authors' knowledge, however, solutions to this problem exist only for vertex-to-vertex \ul{distance} distributions \cite{SadilekJCEM18}, with the case of vertex angles never addressed before.
Since the focus of this article is to develop \ac{ISAC} estimators, we leave this matter for a future contribution and meanwhile consider the proposed highly-accurate approximate model, as illustrated in Figure \ref{fig:angle_dist_PDF}.}\,\!\! \acp{PDF} of the scattering angles $\theta$ can be obtained by evaluating equation \eqref{eq:scatter_angle} for all possible combinations of admissible locations of \ac{UE}, \ac{AP} and voxel within the \ac{ROI}, respectively given by $\mathbf{c}_{U}$, $\mathbf{c}_{A}$ and $\mathbf{c}_{V}$, and examples of the latter for an environment voxelated at various resolutions are shown in Fig. \ref{fig:angle_dist_PDF}.

It is visible in Figure \ref{fig:angle_dist_PDF} that for a sufficiently large  $N_V$, the distribution of scattering angles $\theta$ can be well modeled by a mixture of two scaled beta distributions, namely, ${f}_{\mathrm{x}}(x) = \gamma \!\cdot\! \beta(a_1, b_1) + (1-\gamma) \!\cdot\! \beta(a_2, b_2)$, with support $x \in [0,\pi]$ and where $\gamma$ is a weighing factor and the quantities $a_1, a_2, b_1, b_2$ are shape parameters optimised to match the empirical data obtained by evaluating equation \eqref{eq:scatter_angle}, with $\mathbf{c}_{U}$, $\mathbf{c}_{A}$ and $\mathbf{c}_{V}$ taken randomly within the voxelated grid.

Utilizing this empirical stochastic-geometric approach, the increasingly popular voxelated model utilized in various related works \cite{Tong_JSTSP21,Tao_WCSP20, Zhang_ICC2022, MoralesTITS2017, Pirker_ICCVW11, Collins_MCCA07} can be improved by the incorporation of random blockages of \ac{NLOS} subpaths, in proportion to the complement cumulative distribution of the approximated beta mixture \ac{PDF}, and in accordance to the scattering angles at each voxel, as a function of a selected critical angle.


\vspace{-2ex}
\subsection{Signal Model}
\label{sec:signal_model}
\vspace{-0.5ex}

Consider an uplink communication scenario between the group of $N_U$ \acp{UE} and a total of $N_A$ \acp{AP}, under the models described above in Subsections \ref{sec:voxelation_model} through \ref{sec:environment_model}, with the $N_A$ \acp{AP} connected to a \ac{CPU} via error-free fronthaul links of unlimited throughput, such that the received signals at all $N_A N_R$ receive antennas are aggregated without loss of information or delay.

Then, the aggregated received signal matrix $\mathbf{Y}$, over $N_T$ discrete transmission instances (symbol slots) is given by
\vspace{-1ex}
\begin{equation}
\mathbf{Y} = \mathbf{G} \mathbf{X} + \mathbf{W} \in \mathbb{C}^{N_A N_R \times N_T},
\vspace{-1.5ex}
\label{eq:received_signal}
\end{equation}
where $\mathbf{G} \in \mathbb{C}^{N_A N_R \times N_U}$ is the effective channel matrix as described in Subsection \ref{sec:channel_model}; $\mathbf{X} \in \mathbb{C}^{N_U \times N_T}$ is the transmit signal matrix collecting the symbols from all $N_U$ \acp{UE}, each drawn from the constellation $\mathcal{X}$ {of} cardinality $N_{\mathcal{X}}$; and $\mathbf{W} \in \mathbb{C}^{N_A N_R \times N_T}$ is the receive \ac{AWGN} matrix with \ac{i.i.d.} elements drawn from $\mathcal{CN}(0, N_0)${,} where $N_0$ is the noise variance.

The transmit signal $\mathbf{X}$ comprises of a pilot block $\mathbf{X}_P \in \mathbb{C}^{N_U \times N_P}$ and a data block $\mathbf{X}_D \in \mathbb{C}^{N_U \times N_D}$, such that
\vspace{-1.5ex}
\begin{equation}
\begin{matrix}\mathbf{X}\end{matrix} = \begin{bmatrix} \mathbf{X}_P & \! \mathbf{X}_D \end{bmatrix} \begin{matrix}\in \mathbb{C}^{N_U \times N_T}, \end{matrix}
\label{eq:transmit_signal}
\vspace{-1ex}
\end{equation}
where $N_P$ and $N_D$ denote the number of symbol slots allocated to the pilot and data sequences, respectively, with $N_T = N_P + N_D$; and where the pilot symbol matrix $\mathbf{X}_P$ is assumed to be perfectly known at the \ac{CPU}.

In view of equations \eqref{eq:effective_channel}, \eqref{eq:received_signal} and \eqref{eq:transmit_signal}, the goal of the \ac{ISAC} schemes to be hereafter presented can be concisely stated.
The communication objective of the \ac{CPU} is to estimate the unknown data symbol matrix $\mathbf{X}_D$, under the knowledge of only the pilot symbols in $\mathbf{X}_P$, after the estimation of the channel matrix $\mathbf{G}$.
In turn, the sensing objective is to extract the voxelated model of the environment as the vector of occupancy coefficients $\mathbf{v}$, from the said channel matrix $\mathbf{G}$.

\vspace{-1ex}
\section{Proposed \ac{ISAC} Solution}
\label{sec:proposed_solution}

By combining the channel decomposition model of equation \eqref{eq:effective_channel}, the received signal model of equation \eqref{eq:received_signal}, and the transmit signal in model of equation \eqref{eq:transmit_signal}, the overall system model becomes
\vspace{-1.5ex}
\begin{equation}
\mathbf{Y} = \overbrace{\big(\mathbf{H} + \mathbf{A} \diag{\mathbf{v}}\!\mathbf{B}\big)}^{\triangleq \mathbf{G}}  \overbrace{\big[\mathbf{X}_P \; \mathbf{X}_D\big]}^{\triangleq \mathbf{X}}+ \mathbf{W}  \in \mathbb{C}^{N_A N_R \times N_T},
\label{eq:full_system}
\vspace{-1.5ex}
\end{equation}
where the unknown variables of interest are the environment (voxel coefficients) vector $\mathbf{v}$ and the data symbol matrix $\mathbf{X}_D$.

Similar to related literature \cite{Tong_JSTSP21,Tao_WCSP20, Zhang_ICC2022, MoralesTITS2017, Pirker_ICCVW11, Collins_MCCA07}, it is assumed hereafter that the \ac{LOS} channel $\mathbf{H}$, and the \ac{UE}-to-voxel and voxel-to-\ac{AP} subpath of components $\mathbf{A}$ and $\mathbf{B}$ in equation \eqref{eq:full_system} are known, which still leaves an atypical relationship between the two variables
$\diag{\mathbf{v}}$ and $\mathbf{X}_D$.
In particular, the latter unknowns are related, under equation \eqref{eq:full_system}, by an asymmetric bilinear system, requiring sophisticated algorithms to be either decoupled or jointly estimated \cite{Parker_TSP14,Iimori_TWC21,Iimori_ICC22, Parker_JSTSP16, Yuan_TSP21, Rou_Asilomar22_QSM}.

In light of the above, we propose in the sequel two novel \ac{ISAC} solutions for the joint estimation problem of the asymmetric bilinear system expressed by  \eqref{eq:full_system}, by leveraging the \ac{GaBP} \ac{MP} framework.
The first proposed method incorporates two separate \ul{linear} estimation modules for each of the unknown variables $\mathbf{v}$ and $\mathbf{X}_D$, estimating them in an alternate fashion via feedback between the two modules.
In turn, the second method utilizes only a single \ul{bilinear} estimation module which enables the simultaneous extraction of both unknown variables.

\vspace{-1.5ex}
\subsection{Proposed Alternating Linear ISAC Algorithm (\acs{AL-ISAC})}
\label{sec:ALISAC}

The first proposed method, dubbed the ``\Ac{AL-ISAC}'' algorithm, leverages two separate linear \ac{GaBP} \ac{MP} modules based on equation \eqref{eq:full_system}, which are respectively described as:
\textit{1)} a linear \ac{GaBP} module to estimate the environment vector $\mathbf{v}$,  given the transmit signal $\mathbf{X}$, and conversely; and
\textit{2)} a linear \ac{GaBP} module to estimate the transmit signal matrix $\mathbf{X}$, given the environment vector $\mathbf{v}$.

The two linear \ac{GaBP} modules and the constituting \ac{MP} rules are derived in the next subsections, followed by the construction of the full \ac{ISAC} algorithm encompassing the two derived modules.

\subsubsection{Linear \ac{GaBP} for Environment Vector $\mathbf{v}$} $~$

\begin{figure}[H]
\vspace{-1.5ex}
\centering
\includegraphics[width=0.75\textwidth]{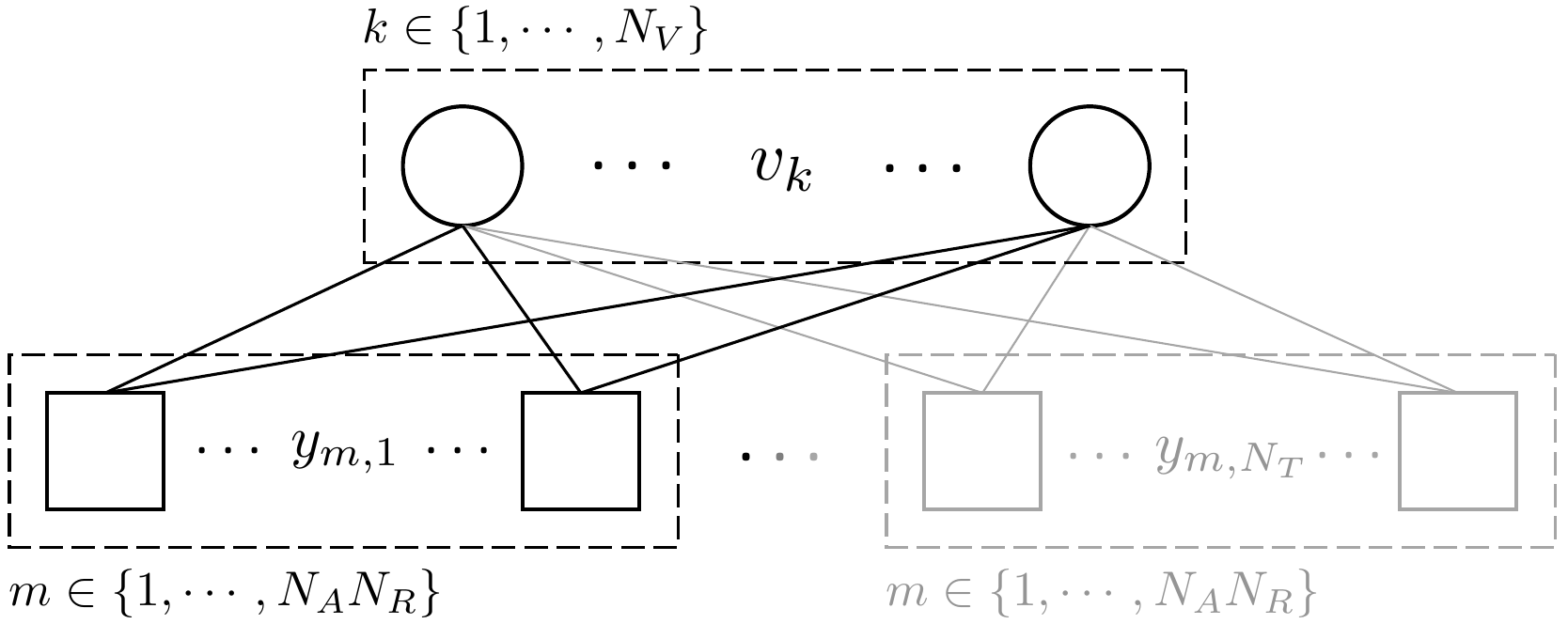}
\vspace{-2ex}
\caption{Factor graph of the linear system formulated for the estimation of $\mathbf{v}$.}
\label{fig:factorgraph_linear_v}
\vspace{-2ex}
\end{figure}

The linear \ac{GaBP} algorithm operates on only one unknown variable, so that in order to estimate $\mathbf{v}$, the entire transmit signal matrix $\mathbf{X}$ must be assumed known, in addition to the known channel matrices $\mathbf{H}, \mathbf{A},$ and $\mathbf{B}$.
Assuming knowledge of $\mathbf{X}$, the system in \eqref{eq:full_system} may be reformulated as 
\vspace{-0.5ex}
\begin{equation}
\mathbf{Y} = \mathbf{H}\mathbf{X} + \mathbf{A} \diag{\mathbf{v}} \!\mathbf{B} \mathbf{X} + \mathbf{W} \in \mathbb{C}^{N_A N_R \times N_T},
\label{eq:full_system_for_v}
\vspace{-0.5ex}
\end{equation}
where, since the channel matrix $\mathbf{A} \in \mathbb{C}^{N_A N_R \times N_V}$ and the matrix products $\mathbf{H}\mathbf{X} \in \mathbb{C}^{N_A N_R \times N_T}$ and $\mathbf{B} \mathbf{X} \in \mathbb{C}^{N_V \times N_T}$ and are known, the described system in \eqref{eq:full_system_for_v} is linear on $\mathbf{v}$, to which a corresponding factor graph may be obtained as in Fig. \ref{fig:factorgraph_linear_v}.

Each element $y_{m,t}$ of the receive signal of $\mathbf{Y}$, with $m \in \{1,\cdots,N_AN_R\}$ and $t \in \{1, \cdots, N_T\}$, corresponds to the factor nodes (square nodes) and each element $v_k$ of the unknown environment variable $\mathbf{v}$, with $k \in \{1,\cdots,N_V\}$, correspond to the variable nodes (circular nodes).
In turn, each ($m,t$)-th factor node on the factor graph has a corresponding soft-replica of each variable node element $v_k$ denoted by $\hat{v}_{k:m,t}$, with the corresponding \ac{MSE} given by
\vspace{-0.5ex}
\begin{equation}
\psi^{v}_{k:m,t} = \Exp_{\mathsf{v}_k}[|v_{k} - \hat{v}_{k:m,t}|^2].
\label{eq:linear_gabp_v_mse}
\vspace{-0.5ex}
\end{equation}

Utilizing the soft-replicas and their \acp{MSE}, the factor nodes perform soft-\ac{IC} on the received signal $y_{m,t}$ for each variable $v_k$, yielding the \ac{IC} symbol
\vspace{-1ex}
\begin{equation}
\bar{y}_{k:m,t}^{v}\! =\! y_{m,t} -\! \sum_{u = 1}^{N_U}\! h_{m,u} x_{u,t} -\! \sum_{i \neq k }^{N_V} \!\Big[ a_{m,i} \hat{v}_{i:m,t}\! \overbrace{\sum_{ u = 1 }^{N_U} b_{i,u} x_{u,t}}^{\triangleq c_{i,t}} \!\Big]\!\! = \! v_{k} \big( a_{m,k} c_{k,t} \big) + \! \overbrace{\sum_{i \neq k }^{N_V}\!  a_{m,i} (v_{i}\! -\! \hat{v}_{i:m,t}) c_{i,t} \!+\! w_{m,t}}^{\text{\acf{SGA}}}, \nonumber
\end{equation}
\vspace{-5.5ex}
\begin{equation}
\label{eq:linear_gabp_v_sic}
\vspace{-2.2ex}
\end{equation}
where $x_{n,t}$ and $w_{m,t}$ are the ($n,t$)-th and ($m,t$)-th elements of $\mathbf{X}$ and $\mathbf{W}$, with $n \in \{1,\cdots,N_U\}$; and $c_{k,t} \triangleq \sum_{ u = 1 }^{N_U} b_{k,u} x_{u,t}$ represents the aggregated incident signal at the $k$-th voxel from all \acp{UE}.

Next, by leveraging the \ac{CLT}, the sum of difference and noise terms are approximated as a complex Gaussian scalar, such that the \ac{PDF} of the interference-cancelled symbols $\bar{y}_{k:m,t}^{v}$ can be modeled as
\vspace{-1ex}
\begin{equation}
\mathbb{P}_{\bar{\mathsf{y}}^{\mathsf{v}}_{k:m,t}}(\bar{y}^{v}_{k:m,t}|v_{k}) \propto \mathrm{exp} \Big[ \!-\!\frac{|\bar{y}^{v}_{k:m,t} -  v_k  \big( a_{m,k} c_{k,t} \big) |^2}{\nu^{v}_{k:m,t}} \Big], 
\label{eq:linear_gabp_v_cpdf}
\end{equation}
with the corresponding conditional variance $\nu^v_{k:m,t}$ given by
\vspace{-1ex}
\begin{equation}
\nu^v_{k:m,t} = \Exp_{\mathsf{v}_k}[|\bar{y}_{k:m,t}^{v} - (a_{m,k} c_{k,t}) v_{k} |^2] = \sum_{i \neq k}^{N_V} |a_{m,i}|^2 |c_{i,t}|^2 \psi^v_{i:m,t} + N_0,
\vspace{-1ex}
\label{eq:linear_gabp_v_cvar}
\end{equation}
where $N_0\triangleq\Exp_{\mathsf{w}_{m,t}}[|w_{m,t}|^2]$ is the noise variance.

The conditional variances for all $v_k$ are computed by all factor nodes, and the message is sent to the corresponding variable nodes.
Consequently, the $k$-th variable node obtains the $N_A N_R N_T$ conditional variances from all factor nodes, from which the extrinsic belief $\ell^v_k$ is computed.
In \ac{GaBP}, self-interference is suppressed by excluding the conditional \ac{PDF} of $\bar{y}_{k:m,t}$ at the $k$-th variable node to yield $\ell^v_{k:m,t}$ with the \ac{PDF}
\begin{equation}
\mathbb{P}_{\mathsf{l}^{\mathsf{v}}_{k:m,t}}(\ell^v_{k:m,t} | v_k) = \! \! \! \! \prod_{p \neq m}^{N_A N_R} \! \prod_{ q \neq t}^{N_T} \mathbb{P}_{\bar{\mathsf{y}}^{\mathsf{v}}_{k:m,t}}(\bar{y}^{v}_{k:p,q}|v_{k}) \propto \mathrm{exp} \Bigg[ -\frac{|v_k - \mu^v_{k:m,t}|^2}{\varPsi^v_{k:m,t}} \Bigg],\\
\label{eq:linear_gabp_v_expdf}
\end{equation}
where the extrinsic mean $\mu^v_{k:m,t}$ and variance $\varPsi^v_{k:m,t}$ is respectively given by
\begin{equation}
\mu^v_{k:m,t} = \varPsi^v_{k:m,t} \cdot \Bigg(\sum_{p \neq m}^{N_A N_R}\!\!\sum_{q \neq t}^{N_T} \!\frac{ \big(a_{p,k} c_{k,t}\big)^* \cdot \bar{y}^{v}_{k:p,q}}{ \nu^{v}_{k:p,q}}\Bigg)
~\text{and}~
\varPsi^v_{k:m,t} = \Bigg(\sum_{p \neq m}^{N_A N_R} \!\!\sum_{q \neq t}^{N_T} \!\frac{|a_{p,k}|^2 |c_{k,t} |^2}{ \nu^{v}_{k:p,q}} \Bigg)^{\!-1}\!\!\!\!.
\label{eq:linear_gabp_v_exmean_and_var}
\end{equation}

Finally, by following the Bayes rule, the updated posterior may be obtained by combining the \ac{PDF} of the extrinsic belief and the prior distribution of $v_k$, from which the updated soft-replica is obtained as
\vspace{-0.5ex}
\begin{equation}
\hat{v}_{k:m,t} =  \dfrac{ \Exp_{\mathsf{v}_k}\!\!\left[\mathbb{P}_{\mathsf{l}^{\mathsf{v}}_{k:m,t}}(\ell^v_{k:m,t} | v_k) \cdot \mathbb{P}_{\mathsf{v}_k}(v_k) \right]}{\int_{\mathsf{v_k}} \!\! \mathbb{P}_{\mathsf{l}^{\mathsf{v}}_{k:m,t}}(\ell^v_{k:m,t} | v_k) \cdot \mathbb{P}_{\mathsf{v}_k}(v_k)},
\label{eq:linear_gabp_v_updatedsr}
\end{equation}
where the normalizing factor in the denominator is the integrated updated posterior over $\mathbb{C}$.

Similarly, the updated error variance of the soft-replica is obtained by evaluating
\vspace{-1ex}
\begin{equation}
\psi^v_{k:m,t} =  \dfrac{ \Var_{\mathsf{v}_k}\!\!\left[\mathbb{P}_{\mathsf{l}_{k:m,t}^{\mathsf{v}}}(\ell^v_{k:m,t} | v_k) \cdot \mathbb{P}_{\mathsf{v}_k}(v_k) \right]}{\int_{\mathsf{v_k}} \!\! \mathbb{P}_{\mathsf{l}^{\mathsf{v}}_{k:m,t}}(\ell^v_{k:m,t} | v_k) \cdot \mathbb{P}_{\mathsf{v}_k}(v_k)} .
\label{eq:linear_gabp_v_updatedmse}
\end{equation}

Given the information of the voxel coefficient distributions, \textit{i.e.,} binary coefficients with a discrete prior given by a Bernoulli distribution with occupancy probability $E_v \triangleq \mathbb{P}_{\mathsf{v_k}}(v_k=1)$, the soft-replica and its \ac{MSE} can be efficiently obtained in closed-form, respectively given by
\begin{eqnarray}
\label{eq:linear_gabp_v_updatedsr_discrete}
&\hat{v}_{k:m,t}  = \bigg(1 + \dfrac{1 - E_v}{E_v} \mathrm{exp}\Big( -\dfrac{|{\mu}^v_{k:m,t}|^2 - |1 - {\mu}^v_{k:m,t}|^2}{\varPsi^{v}_{k:m,t}} \Big)\bigg)^{\!-1}\!\!\!\!,&\\
\label{eq:linear_gabp_v_updatedmse_discrete}
&\psi^v_{k:m,t} = (\hat{v}_{k:m,t})^2 + E_v - 2E_v\hat{v}_{k:m,t}.&
\end{eqnarray}

The updated soft-replica and the \ac{MSE} of each variable node are then transmitted back to all factor nodes for the next iteration of the \ac{GaBP} \ac{MP} algorithm.
After a given number of \ac{GaBP} iterations to refine the soft-estimates, a belief consensus is taken at each variable node across the soft-replicas to obtain a single estimate $\tilde{\mathbf{v}}$ by
\begin{equation}
\mathbb{P}_{\tilde{\mathsf{l}}^{\mathsf{v}}_{k}}(\tilde{\ell}^v_{k} | v_k) = \! \! \! \! \prod_{p = 1}^{N_A N_R} \!\prod_{q = 1}^{N_T} \mathbb{P}_{\bar{\mathsf{y}}^{\mathsf{v}}_{k:m,t}}(\bar{y}^{v}_{k:p,q}|v_{k}) \propto \mathrm{exp} \left[ -\frac{|v_k - \tilde{\mu}^v_{k}|^2}{\tilde{\varPsi}^v_{k}} \right],\\
\label{eq:linear_gabp_v_conpdf}
\end{equation}
with consensus mean $\tilde{\mu}^v_{k}$ and variance $\tilde{\varPsi}^v_{k}$ expressed as
\begin{equation}
\tilde{\mu}^v_{k} = \tilde{\varPsi}^v_{k} \cdot \Bigg(\sum_{p =1}^{N_A N_R} \!\sum_{q =1}^{N_T} \frac{ \big(a_{p,k} c_{k,t}\big)^* \cdot \bar{y}^{v}_{k:p,q}}{ \nu^{v}_{k:p,q}}\Bigg)
\quad \text{and} \quad
\tilde{\varPsi}^v_{k} = \Bigg(\sum_{p =1}^{N_A N_R} \!\sum_{q =1}^{N_T} \frac{|a_{p,k}|^2 |c_{k,t}|^2}{ \nu^{v}_{k:p,q}} \Bigg)^{\!\!\!-1}\!\!\!\!,
\label{eq:linear_gabp_v_conmean_and_var}
\end{equation}
which is consequently used to yield the final estimate by
\begin{equation}
\tilde{v}_{k} = \dfrac{\Exp_{{\mathsf{v}}_k}\!\Big[ \mathbb{P}_{\tilde{\mathsf{l}}^{\mathsf{v}}_{k}}(\tilde{\ell}^v_{k} | v_k) \cdot \mathbb{P}_{{\mathsf{v}}_k}(v_k)\Big]}{\int_{{v}_k} \!\! \mathbb{P}_{\tilde{\mathsf{l}}^{\mathsf{v}}_{k}}(\tilde{\ell}^v_{k} | v_k) \cdot \mathbb{P}_{{\mathsf{v}}_k}(v_k)}.
\label{eq:linear_gabp_v_final_estimate}
\end{equation}

The above \ac{MP} equations \eqref{eq:linear_gabp_v_mse} to \eqref{eq:linear_gabp_v_final_estimate} fully describe the linear \ac{GaBP} module to estimate the voxel environment $\mathbf{v}$, which is summarized as a pseudocode in Algorithm \ref{alg:linear_gabp_v}.
It is important to note that the signal matrix $\mathbf{X}$ is assumed \ul{given} -- \textit{i.e.,} $\mathbf{X}$ is \ul{not} estimated by Algorithm \ref{alg:linear_gabp_v} -- and therefore is kept constant throughout the iterations, as seen by the pre-computation of the effective signals $c_{k,t}$. The complementary block dedicated to the estimation of $\mathbf{X}$ given $\mathbf{v}$ will be described subsequently, leading to an alternate approach as previously mentioned.
The algorithm also contains a damping update mechanism \cite{Som_ITW10} using a damping factor $\eta \in [0,1]$ to prevent early convergence to a local optimum.

\begin{algorithm}[H]
\hrulefill
\begin{algorithmic}[1] 
\vspace{-1.5ex}
\setlength{\baselineskip}{18pt}%
\Statex \hspace{-3ex} {\bf{Inputs:}} {Received signal matrix $\mathbf{Y}$, channel matrices  $\mathbf{H}$, $\mathbf{A}$, and $\mathbf{B}$, transmit signal matrix $\mathbf{X}$, noise variance $N_0$, prior distribution of voxels $\mathbb{P}_{\mathsf{v}_k}(v_k)$, and pre-computation of $\mathbf{C} = \mathbf{B}\mathbf{X}$. \vspace{-0.5ex}}
\Statex \hspace{-3ex} {\bf{Outputs:}} Estimated voxel environment vector $\tilde{\mathbf{v}}$. \vspace{-2.5ex}
\Statex \hspace{-4ex}\hrulefill \vspace{-1ex}
\State Initialize the soft-replica at all variable nodes as $\hat{v}_{k:m,t} = \Exp_{\mathsf{v}_k}[v_{k}]$;
\Comment{$\forall k,m,t$}
\State Initialize the \acp{MSE} at all variable nodes $\psi^v_{k:m,t}$ via \eqref{eq:linear_gabp_v_mse}; \Comment{$\forall k,m,t$}
\Statex \hspace{-1ex} \textbf{Until} termination criteria is satisfied$\color{black}^{*}$ \textbf{do} 
\State \hspace{1ex} Compute the soft-\ac{IC} performed received signal $\bar{y}_{k:m,t}^{v}$ via \eqref{eq:linear_gabp_v_sic}; \Comment{$\forall k,m,t$}
\State \hspace{1ex} Compute the conditional variance ${\nu}^{v}_{k:m,t}$ via \eqref{eq:linear_gabp_v_cvar}; \Comment{$\forall k,m,t$}
\State \hspace{1ex} Compute the extrinsic mean ${\mu}^{v}_{k:m,t}$ and variance ${\varPsi}^{v}_{k:m,t}$ via \eqref{eq:linear_gabp_v_exmean_and_var}; \Comment{$\forall k,m,t$}
\State \hspace{1ex} Compute the new soft-replica $\hat{v}_{k:m,t}$ and the \ac{MSE} $\psi^v_{k:m,t}$ via \eqref{eq:linear_gabp_v_updatedsr}-\eqref{eq:linear_gabp_v_updatedmse}; \Comment{$\forall k,m,t$}
\State \hspace{1ex} Update the soft-replica $\hat{v}_{k:m,t}$ and the \ac{MSE} $\psi^v_{k:m,t}$ via damping \cite{Som_ITW10}; \Comment{$\forall k,m,t$}
\Statex  \textbf{end}
\State Compute the consensus mean $\tilde{\mu}^v_{k}$ and variance $\tilde{\varPsi}^v_{k}$ via \eqref{eq:linear_gabp_v_conmean_and_var}; \Comment{$\forall k$}
\State Compute the final soft-estimate $\tilde{v}_{k}$ via \eqref{eq:linear_gabp_v_final_estimate}; \Comment{$\forall k$}
\caption[]{: Linear \ac{GaBP} Estimator for Environment (Voxel Coefficients) Vector $\mathbf{v}$}
\label{alg:linear_gabp_v}
\end{algorithmic}
\end{algorithm}
\vspace{-4.5ex}
{\footnotesize \noindent \color{black}$^{*}\!$ The termination criteria can be set as the maximum number of \ac{MP} iterations or the convergence threshold of the soft-replicas, \\[-2.2ex] depending on the desired accuracy or complexity. A discussion o appropriate values are provided in Section IV-C.}

\vspace{1ex}
\subsubsection{Linear \ac{GaBP} for Signal Matrix $\mathbf{X}$} $~$

\begin{figure}[b]
\vspace{-5ex}
\centering
\includegraphics[width=0.7\textwidth]{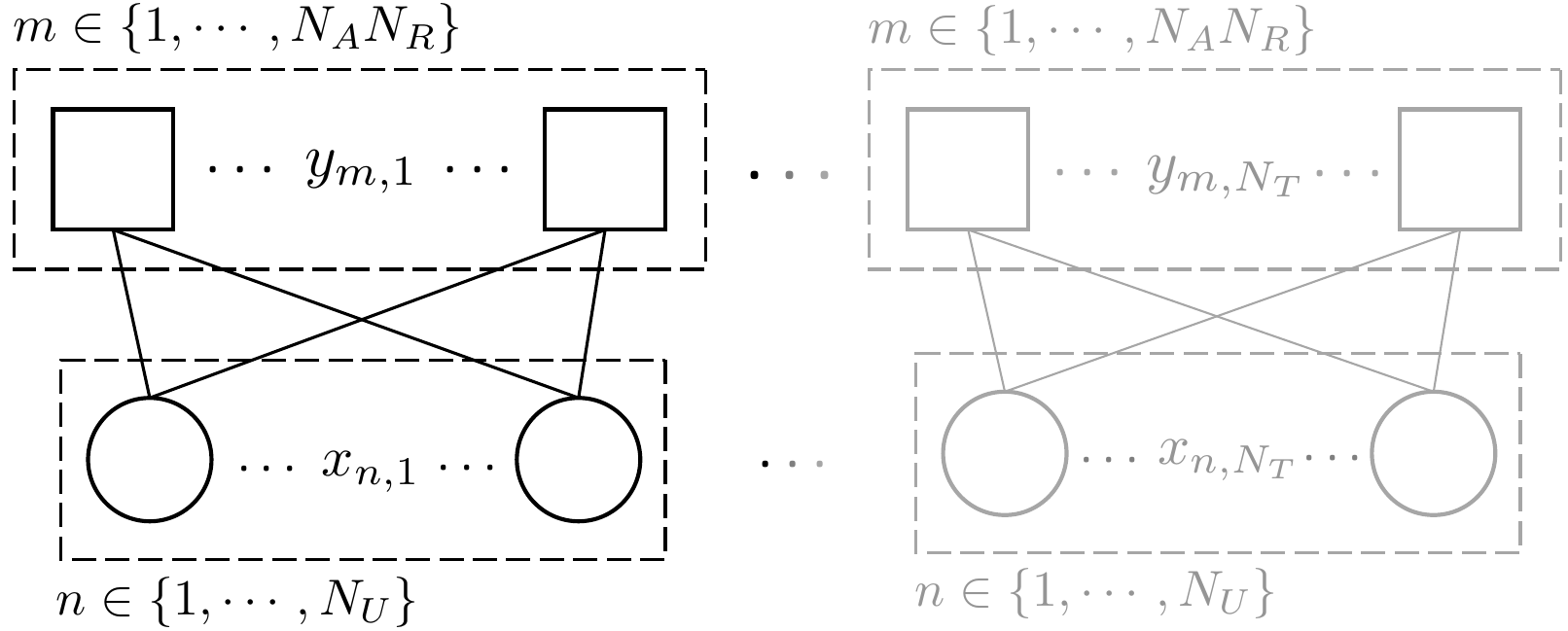}
\vspace{-2ex}
\caption{The separated factor graphs of the linear system for $\mathbf{X}$.}
\label{fig:factorgraph_linear_x}
\vspace{-2ex}
\end{figure}

In order to derive the linear \ac{GaBP} module to estimate the signal matrix $\mathbf{X}$ given $\mathbf{v}$, the system model in \eqref{eq:full_system} is first reduced to the one in \eqref{eq:received_signal}, with the effective channel $\mathbf{G} \triangleq \big(\mathbf{H} + \mathbf{A}\diag{\mathbf{v}}\!\mathbf{B}\big)$, with the corresponding factor graph as illustrated in Fig. \ref{fig:factorgraph_linear_x}, where each element $x_{n,t}$ of the unknown signal matrix $\mathbf{X}$ with $n \in \{1,\cdots,N_U\}$, is the variable node (circular nodes).

Notice that in this case, since the variable $\mathbf{X}$ is two-dimensional, the linear system results in a factor graph that is separated into ``pages'', such that the variable nodes and factor nodes corresponding to different $t \in \{1,\cdot,N_T\}$ are independent and that the messages are only exchanged by nodes with the same time index $t$. 
Other than this separation of factor graphs, the derivation of the \ac{MP} rules is similar to that of Algorithm \ref{alg:linear_gabp_v}.

The soft-replica of the transmit signal matrix element $x_{n,t}$ to the $(m,t$)-th factor node is denoted by $\hat{x}_{m,t:n,t}$, with the corresponding \ac{MSE} given by 
\vspace{-0.5ex}
\begin{equation}
\psi^x_{m,t:n,t} = \Exp_{\mathsf{x}_{n,t}}[|x_{n,t} - \hat{x}_{m,t:n,t}|^2].
\label{eq:linear_gabp_x_mse}
\vspace{-0.5ex}
\end{equation} 

The soft-replica and the \ac{MSE} are used in the soft-\ac{IC} of the received signals for $x_{n,t}$, following
\vspace{-2.5ex}
\begin{equation}
\bar{y}^{x}_{n,t:m,t} = y_{m,t} - \sum^{N_U}_{u \neq n} g_{m,u} \hat{x}_{u,t:m,t} = g_{m,n} x_{n,t} + \sum^{N_U}_{u \neq n} g_{m,u} (x_{u,t} - \hat{x}_{u,t:m,t}) + w_{m,t},
\label{eq:linear_gabp_x_sic}
\end{equation}
with the conditional \ac{PDF} described by 
\vspace{-0.5ex}
\begin{equation}
\mathbb{P}_{\bar{\mathsf{y}}^{\mathsf{x}}_{n,t:m,t}}(\bar{y}^{x}_{n,t:m,t}|x_{n,t}) \propto \mathrm{exp} \left[ -\frac{|\bar{y}^{x}_{n,t:m,t} -  g_{m,n} x_{n,t} |^2}{\nu^{x}_{n,t:m,t}} \right], 
\label{eq:linear_gabp_x_cpdf}
\end{equation}
and the conditional variance $\nu^x_{n,t:m,t}$ is given by
\vspace{-1.5ex}
\begin{equation}
\nu^x_{n,t:m,t} = \Exp_{\mathsf{x}_{n,t}}[|\bar{y}^{x}_{n,t:m,t} -  g_{m,n} x_{n,t} |^2] = \sum_{u \neq n}^{N_U} |g_{m,u}|^2 \psi^x_{u,t:m,t} + N_0.
\label{eq:linear_gabp_x_cvar}
\vspace{-1ex}
\end{equation}

The conditional \acp{PDF} are combined with self-interference cancellation at the variable nodes to yield the extrinsic beliefs $\ell^x_{n,t:m,t}$ following
\begin{equation}
\mathbb{P}_{\mathsf{l}^{\mathsf{x}}_{n,t:m,t}}(\ell^x_{n,t:m,t} | x_{n,t}) =  \prod_{p \neq m}^{N_A N_R} \mathbb{P}_{\bar{\mathsf{y}}^{\mathsf{x}}_{n,t:p,t}}(\bar{y}^{x}_{n,t:p,t}|x_{n,t}) \propto \mathrm{exp} \left[ -\frac{|x_{n,t} - \mu^x_{n,t:m,t}|^2}{\varPsi^x_{n,t:m,t}} \right],\\
\label{eq:linear_gabp_x_expdf}
\end{equation}
with the extrinsic mean $\mu^x_{n,t:m,t}$ and variance $\varPsi^x_{n,t:m,t}$  respectively given by
\begin{equation}
\mu^x_{n,t:m,t} = \varPsi^x_{n,t:m,t} \cdot \Bigg(\mathlarger\sum_{p \neq m}^{N_A N_R} \frac{ (g_{p,n})^* \! \cdot \bar{y}^{x}_{n,t:p,t}}{ \nu^{x}_{n,t:p,t}}\Bigg)
\quad \text{and} \quad
\varPsi^x_{n,t:m,t} = \Bigg(\mathlarger\sum_{p \neq m}^{N_A N_R} \frac{|g_{p,n}|^2}{ \nu^{x}_{n,t:p,t}} \Bigg)^{\!-1}\!\!\!\!.
\label{eq:linear_gabp_x_exvar_and_mean}
\end{equation}

In turn, since the symbols have a uniformly discrete prior from the symbol constellation $\mathcal{X}$, with the symbol probability $\mathbb{P}_{\mathsf{x}}(x) = 1\slash|\mathcal{X}|$, their soft-replicas and \acp{MSE} are obtained by
\begin{eqnarray}
&\hat{x}_{n,t:m,t} = \dfrac{ \sum_{x \in \mathcal{X}} x \cdot \mathbb{P}_{\mathsf{l}^{\mathsf{x}}_{n,t:m,t}}(\ell^x_{n,t:m,t} | x) \cdot \mathbb{P}_{\mathsf{x}}(x)}{\sum_{x \in \mathcal{X}} \mathbb{P}_{\mathsf{l}^{\mathsf{x}}_{n,t:m,t}}(\ell^x_{n,t:m,t} | x) \cdot \mathbb{P}_{\mathsf{x}}(x)} \;,&
\label{eq:linear_gabp_x_updatedsr}\\[1ex]
&\psi^x_{n,t:m,t} = \dfrac{ \sum_{x \in \mathcal{X}} x^2 \cdot \mathbb{P}_{\mathsf{l}^{\mathsf{x}}_{n,t:m,t}}(\ell^x_{n,t:m,t} | x) \cdot \mathbb{P}_{\mathsf{x}}(x)}{\sum_{x \in \mathcal{X}} \mathbb{P}_{\mathsf{l}^{\mathsf{x}}_{n,t:m,t}}(\ell^x_{n,t:m,t} | x) \cdot \mathbb{P}_{\mathsf{x}}(x)} - (\hat{x}_{n,t:m,t})^2.&
\label{eq:linear_gabp_x_updatedmse}
\end{eqnarray}

For the particular case of \ac{MQAM} with $M = 4$, the soft-replica and \ac{MSE} computations reduce to efficient closed-form expressions given by
\begin{equation}
\hat{x}_{n,t:m,t} =  \sqrt{\frac{E_\mathcal{X}}{{2}}} \cdot \left( \mathrm{tanh}\left[ \sqrt{\frac{2}{E_\mathcal{X}}}\cdot \frac{\Re{\{ \mu^x_{n,t:m,t} \}}}{ \varPsi^x_{n,t:m,t}} \right] + j \mathrm{tanh}\left[ \sqrt{\frac{2}{E_\mathcal{X}}}\cdot\frac{\Im{\{ \mu^x_{n,t:m,t} \}}}{\varPsi^x_{n,t:m,t}} \right]\right),
\label{eq:linear_gabp_x_updatedsr_4qam}
\end{equation}
\vspace{-3ex}
\begin{equation}
\psi^{x}_{n,t:m,t} =  E_\mathcal{X} - |\hat{x}_{n,t:m,t}|^2,
\label{eq:linear_gabp_x_updatedmse_4qam}
\vspace{-1ex}
\end{equation}
where $E_\mathcal{X} \triangleq \Exp_{x}[|x|^2]$ denotes the average symbol power of the constellation $\mathcal{X}$, and $\mathrm{tanh}(\cdot)$ denotes the trigonometric hyperbolic tangent function.

Finally, the consensus \ac{PDF}, which is taken after the iterations is given by
\begin{equation}
\mathbb{P}_{\tilde{\mathsf{l}}^{\mathsf{x}}_{n,t}}(\tilde{\ell}^x_{n,t} | x_{n,t}) = \prod_{p = 1}^{N_A N_R} \mathbb{P}_{\bar{\mathsf{y}}^{\mathsf{x}}_{n,t:m,t}}(\bar{y}^{x}_{n,t:p,t}|x_{n,t}) \propto \mathrm{exp} \left[ -\frac{|x_{n,t} - \tilde{\mu}^x_{n,t}|^2}{\tilde{\varPsi}^x_{n,t}} \right],\\[-0.5ex]
\label{eq:linear_gabp_x_conpdf}
\end{equation}
with the consensus mean $\tilde{\mu}^x_{n,t}$ and variance $\tilde{\varPsi}^x_{n,t}$ expressed as
\begin{equation}
\tilde{\mu}^x_{n,t} = \varPsi^x_{n,t} \cdot \Bigg(\mathlarger\sum_{p = 1}^{N_A N_R} \frac{ (g_{p,n})^* \cdot \bar{y}^{x}_{n,t:p,t}}{ \nu^{x}_{n,t:p,t}}\Bigg)
\quad \text{and} \quad
\tilde{\varPsi}^x_{n,t} = \Bigg(\mathlarger\sum_{p = 1}^{N_A N_R} \frac{|g_{p,n}|^2}{ \nu^{x}_{n,t:p,t}}\Bigg)^{\!\!\!-1}\!\!\!\!\!,
\label{eq:linear_gabp_x_conmean_and_var}
\end{equation}
yielding the final soft estimate \vspace{-1ex}
\begin{equation}
\tilde{x}_{n,t} = \dfrac{ \sum_{x \in \mathcal{X}} x \cdot \mathbb{P}_{\tilde{\mathsf{l}}^{\mathsf{x}}_{n,t}}(\tilde{\ell}^x_{n,t} | x_{n,t}) \cdot \mathbb{P}_{{\mathsf{x}}_{n,t}}(x_{n,t})}{\sum_{x \in \mathcal{X}} \mathbb{P}_{\tilde{\mathsf{l}}^{\mathsf{x}}_{n,t}}(\tilde{\ell}^x_{n,t} | x_{n,t}) \cdot \mathbb{P}_{{\mathsf{x}}_{n,t}}(x_{n,t})}.
\label{eq:linear_gabp_x_final_estimate}
\end{equation}

Equations \eqref{eq:linear_gabp_x_mse} to \eqref{eq:linear_gabp_x_final_estimate}, collected in the form of a pseudocode in Algorithm \ref{alg:linear_gabp_x}, fully describe the linear \ac{GaBP} module for the estimation of the signal matrix $\mathbf{X}$ given the environment vector $\mathbf{v}$, which together with the previously described module for the estimation of $\mathbf{v}$ given $\mathbf{X}$, completes the proposed \acs{AL-ISAC} scheme.
All that remains is to describe the alternating procedure to estimate both unknown variables, which is addressed in the sequel.

\begin{algorithm}[H]
\hrulefill
\begin{algorithmic}[1] 
\vspace{-1.5ex}
\setlength{\baselineskip}{18pt}%
\Statex \hspace{-3ex} {\bf{Inputs:}} Received signal matrix $\mathbf{Y}$, channel matrices $\mathbf{H}$, $\mathbf{A}$, and $\mathbf{B}$, environment vector $\mathbf{v}$, \qquad noise variance $N_0$, and prior distribution of transmit symbols $\mathbb{P}_{\mathsf{x}_{n,t}}(x_{n,t})$. \vspace{-0.5ex}
\Statex \hspace{-3ex} {\bf{Outputs:}} Estimated transmit signal matrix $\tilde{\mathbf{X}}$. \vspace{-2.5ex}
\Statex \hspace{-4ex}\hrulefill \vspace{-1ex}
\State Compute the effective channel matrix $\mathbf{G} \triangleq \mathbf{H} + \mathbf{A}\diag{\mathbf{v}}\!\mathbf{B}$;
\State Initialize the soft-replica at all variable nodes as $\hat{x}_{n,t:m,t} = \Exp_{\mathsf{x}_{n,t}}[x_{n,t}];$ 
\Comment{$\forall m,n,t$}
\State Initialize the \ac{MSE} at all variable nodes as $\psi^x_{n,t:m,t}$ via \eqref{eq:linear_gabp_x_mse}; \Comment{$\forall m,n,t$}
\Statex \hspace{-1ex} \textbf{Until} termination criteria is satisfied{$\color{black}^*$} \textbf{do} 
\State \hspace{1ex} Compute the soft-\ac{IC} performed received signal $\bar{y}_{n,t:m,t}^{x}$ via \eqref{eq:linear_gabp_x_sic}; \Comment{$\forall m,n,t$}
\State \hspace{1ex} Compute the conditional variance ${\nu}^{x}_{n,t:m,t}$ via \eqref{eq:linear_gabp_x_cvar}; \Comment{$\forall m,n,t$}
\State \hspace{1ex} Compute the extrinsic mean ${\mu}^{x}_{n,t:m,t}$ and variance ${\varPsi}^{x}_{n,t:m,t}$ via \eqref{eq:linear_gabp_x_exvar_and_mean}; \Comment{$\forall m,n,t$}
\State \hspace{1ex} Compute the new soft-replica $\hat{x}_{n,t:m,t}$ and the \ac{MSE} $\psi^x_{n,t:m,t}$ via \eqref{eq:linear_gabp_x_updatedsr} and 
\eqref{eq:linear_gabp_x_updatedmse}; \Comment{$\forall m,n,t$}
\State \hspace{1ex} Update the soft-replica $\hat{x}_{n,t:m,t}$ and the \ac{MSE} $\psi^x_{n,t:m,t}$ via damping \cite{Som_ITW10}; \Comment{$\forall m,n,t$}
\Statex \hspace{-2ex} \textbf{end}
\State Compute the consensus mean $\tilde{\mu}^x_{n,t}$ and variance $\tilde{\varPsi}^x_{n,t}$ via \eqref{eq:linear_gabp_x_conmean_and_var}; \Comment{$\forall n,t$}
\State Compute the final soft-estimate $\tilde{x}_{n,t}$ via \eqref{eq:linear_gabp_x_final_estimate}; \Comment{$\forall n,t$}
\State Project $\tilde{x}_{n,t}$ to the symbol constellation $\mathcal{X}$; \Comment{$\forall n,t$}
\State Output projected $\tilde{x}_{n,t}$ as final hard estimate; \Comment{$\forall n,t$}
\caption[]{: Linear \ac{GaBP} Estimator for Signal Matrix $\mathbf{X}$}
\label{alg:linear_gabp_x}
\end{algorithmic}
\end{algorithm}
\vspace{-5ex}
{\footnotesize \noindent \color{black}$^{*}\!$ The termination criteria can be set in accordance to Section IV-C and as described in Algorithm \ref{alg:linear_gabp_v}.}

\subsubsection{Combined Alternating Modular Structure} $~$
\label{sec:A-GaBP}

With the two estimation modules given by Algorithm \ref{alg:linear_gabp_v} and Algorithm \ref{alg:linear_gabp_x}, either of the two variables $\mathbf{v}$ or $\mathbf{X}$ may be estimated, assuming full information of the other variable.
However, the very inherent problem of \ac{ISAC} in equation \eqref{eq:full_system}, is that \ul{neither} of the variables are fully known such that the linear modules may not be directly applied for estimation.

To address the problem, the proposed \ac{AL-ISAC} algorithm successively applies the two linear \ac{GaBP} modules to estimate the two sets of variables. 
To enable this, the received signal is separated into the blocks corresponding to the pilot phase and the data phase, as
\vspace{-1ex}
\begin{subequations}
\begin{align}
\mathbf{Y}_P &= \big(\mathbf{H} + \mathbf{A}\diag{\mathbf{v}}\!\mathbf{B} \big)\mathbf{X}_P + \mathbf{W}_P \in \mathbb{C}^{N_A N_R \times N_P}, \label{eq:received_signal_p}\\[-0.5ex]
\mathbf{Y}_D &= \big(\mathbf{H} + \mathbf{A}\diag{\mathbf{v}}\!\mathbf{B} \big)\mathbf{X}_D + \mathbf{W}_D \in \mathbb{C}^{N_A N_R \times N_D}, \label{eq:received_signal_d}
\end{align}  
\end{subequations}
\vspace{-1ex}
where $\mathbf{Y} \triangleq [\mathbf{Y}_P\; \mathbf{Y}_D]$ and $\mathbf{W} \triangleq [\mathbf{W}_P\;  \mathbf{W}_D]$, as defined with $\mathbf{X} =[\mathbf{X}_P\; \mathbf{X}_D]$.

First, by using only the pilot phase of the system \eqref{eq:received_signal_p}, Algorithm \ref{alg:linear_gabp_v} is applied to estimate the initial environment vector $\tilde{\mathbf{v}}_{init}$ with the pilot block $\mathbf{X}_P$ as known input signal matrix.

\begin{figure}[H]
\centering
\includegraphics[width=0.95\textwidth]{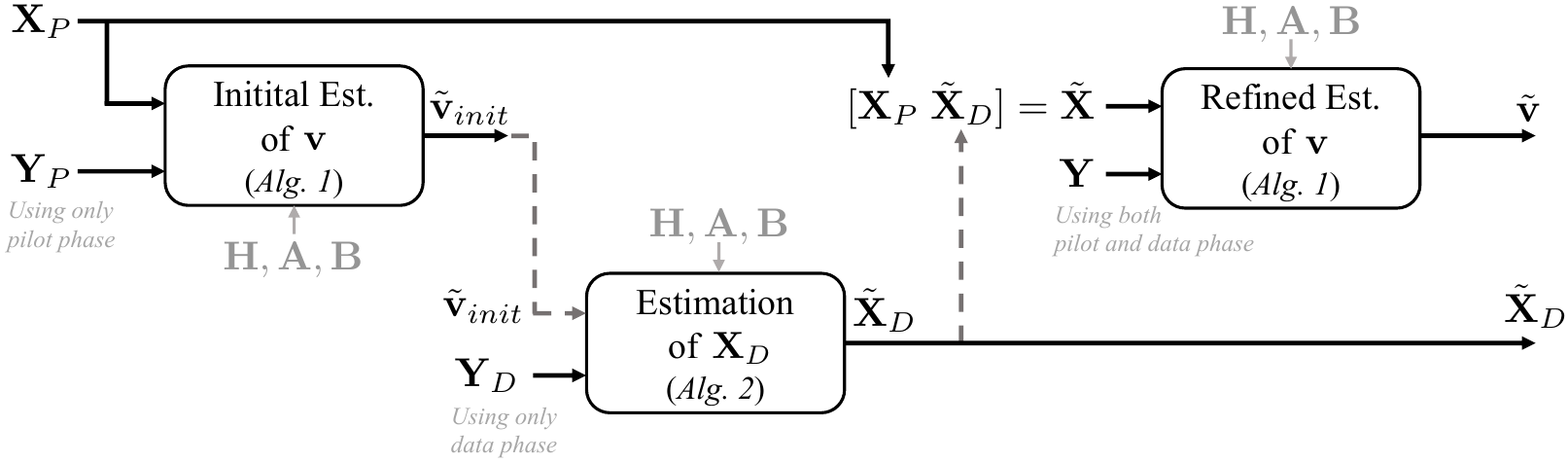}
\vspace{-2ex}
\caption{A schematic diagram of the proposed \ac{AL-ISAC} algorithm {\color{black}(Algorithm \ref{alg:alt_gabp_jcas})}.}
\label{fig:aljcas_schematic}
\vspace{-3.5ex}
\end{figure}

Next, by using the data phase of the system \eqref{eq:received_signal_d}, Algorithm \ref{alg:linear_gabp_x} is applied to estimate the unknown data block $\tilde{\mathbf{X}}_D$ using the initial environment estimate $\tilde{\mathbf{v}}_{init}$ as the known input environment vector.
Finally, the environment vector is obtained by using Algorithm \ref{alg:linear_gabp_x} again, but with the initial environment estimate $\tilde{\mathbf{v}}_{init}$ as the initialization value of the soft-replicas at all factor nodes, and $[\mathbf{X}_P \;\tilde{\mathbf{X}}_D]$ as the input transmit signal matrix.
The described \ac{AL-ISAC} algorithm is illustrated in a schematic form in Fig. \ref{fig:aljcas_schematic}, and summarized as pseudocode in Algorithm \ref{alg:alt_gabp_jcas}.

\vspace{-1ex}
\begin{algorithm}[H]
\hrulefill
\vspace{-1.5ex}
\begin{algorithmic}[1] 
\setlength{\baselineskip}{18pt}%
\Statex \hspace{-3ex} {\bf{Inputs:}} Received signal matrix $\mathbf{Y}$, channel matrices $\mathbf{H}$, $\mathbf{A}$, and $\mathbf{B}$, pilot matrix $\mathbf{X}_P$, noise variance $N_0$, prior distribution of environment and transmit symbols $\mathbb{P}_{\mathsf{v}_{k}}(v_{k})$ and $\mathbb{P}_{\mathsf{x}_{n,t}}(x_{n,t})$. \vspace{-4ex}
\Statex \hspace{-2.2ex} {\bf{Outputs:}} Estimated environment vector $\tilde{\mathbf{v}}$ and estimated {\color{black}data signal matrix $\tilde{\mathbf{X}}_D$}. \vspace{-2.5ex}
\Statex \hspace{-4ex}\hrulefill \vspace{-1ex}
\Statex \vspace{0.5ex}\hspace{-3.5ex} \textit{Using only the block corresponding to the pilot sequence, i.e.,} $t = \{1,\cdots,N_P\}$:
\State Estimate initial environment $\tilde{\mathbf{v}}_{init}$ via Algorithm \ref{alg:linear_gabp_v}, using pilot $\mathbf{X}_P$ as known signal input.
\Statex \vspace{0.5ex}\hspace{-3.5ex} \textit{Using only the block corresponding to the data sequence, i.e.,} $t = \{N_P+1,\cdots,N_P+N_D\}$:
\State Estimate data signal $\tilde{\mathbf{X}}_D$ via Algorithm \ref{alg:linear_gabp_x}, using $\tilde{\mathbf{v}}_{init}$ as known environment input.
\Statex \vspace{0.5ex}\hspace{-3.5ex} \textit{Using the entire time block, i.e.,} $t = \{1,\cdots,N_P+N_D\}$:
\State Estimate $\tilde{\mathbf{v}}$ via Algorithm \ref{alg:linear_gabp_v}, using $[\mathbf{X}_P, \!{\color{black}~\tilde{\mathbf{X}}_D}]$ as the signal input, and $\tilde{\mathbf{v}}_{init}$ as initialization.
\State Output $\tilde{\mathbf{v}}$ as the estimated environment vector, and ${\color{black}\tilde{\mathbf{X}}_D}$ as estimated data signal matrix.
\caption[]{: Proposed \acf{AL-ISAC} Algorithm{\color{black}$^{\dagger}\!$}}
\label{alg:alt_gabp_jcas}
\end{algorithmic}
\end{algorithm}
\vspace{-4.6ex}
{\footnotesize \noindent \color{black}$^{\dagger}\!\!$ Although our simulations indicate that a single iteration (as shown in Fig. \ref{fig:aljcas_schematic}) is sufficient, the linear \ac{GaBP} modules in steps 2 \\[-2.1ex]
and 3 can be iterated multiple times, with feedback at modular level and possibly adaptive \ac{MP} denoising between feedback\\[-2.1ex]
loops. These, and other potential improvements remain open points for a follow up work.
}

Despite having a potential complexity advantage, especially for scenarios with large numbers of \acp{UE}, as shown later in Subsection \ref{sec:complexity}, the alternating approach of the \ac{AL-ISAC} algorithm has the drawback of causing a heavy dependence on the length of the pilot sequence $\mathbf{X}_P$, which as shall be shown in Section \ref{sec:performance}, affects the performance of both environment and data signal estimation, in addition to the obvious trade-off with the total communication throughput.
Aiming to circumvent this deficiency, in the next subsection we propose another \ac{ISAC} method in which both $\mathbf{v}$ and $\mathbf{X}_D$ are estimated simultaneously via a bilinear inference method.

\vspace{-1ex}
\subsection{Proposed Bilinear \ac{ISAC} Algorithm (Bi-ISAC)}
\label{sec:BiGaBP}

In this section, we develop a new \ac{ISAC} algorithm in which the sensing and communication variables $\mathbf{v}$ and $\mathbf{X}_D$ are estimated \ul{in parallel}, by using a bilinear message passing technique which incorporates the uncertainty of both estimates at each iteration, thus requiring only a single estimation module to acquire both variables, as illustrated in Fig. \ref{fig:bijcas_schematic}.

We start by observing that the unique asymmetric bilinear relationship of $\mathbf{v}$ and $\mathbf{X}_D$, as per equation \eqref{eq:full_system_for_v}, prevents the application of recently discovered bilinear estimators, such as the \ac{BiGAMP} \cite{Parker_TSP14}, which operates only on symmetric systems described by equations in the form $\mathbf{Y} = \mathbf{V}\mathbf{X} + \mathbf{W}$ for the joint estimation of the unknowns $\mathbf{V}$ and $\mathbf{X}$; or the parametric \ac{BiGAMP} \cite{Parker_JSTSP16, Yuan_TSP21}, which works on systems with the structure $\mathbf{Y} = \sum_k \! v_k \mathbf{A}_k \mathbf{X} + \mathbf{W}$ to jointly estimate $v_k$ and $\mathbf{X}$ with known $\mathbf{A}_k$.

In contrast to these two examples, the problem dealt with here is, as described by equation \eqref{eq:full_system}, in neither of the aforementioned forms, nor can it be transformed to fit general bilinear forms, which implies that new, purpose-built \ac{BiGaBP} \cite{Iimori_TWC21, Iimori_ICC22, Rou_Asilomar22_QSM} \ac{MP} rules must be derived for its solution. Therefore, the \ac{BiGaBP} message passing is performed on a tripartite factor graph as illustrated in Fig. \ref{fig:factorgraph_bilinear}, where the factor nodes (square

$~$

\vspace{-5ex}
\begin{minipage}[H]{0.34\textwidth}
\vspace{-1ex}
\begin{figure}[H]
\centering
\includegraphics[width=\textwidth]{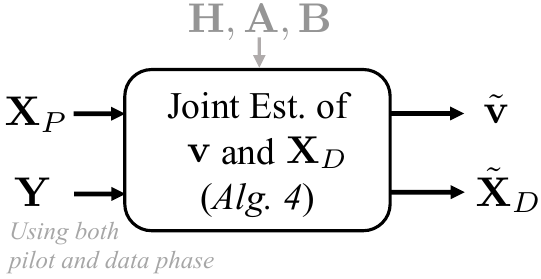}
\caption{A schematic diagram of the proposed \acs{Bi-ISAC} algorithm {\color{black}(Algorithm \ref{alg:bi_gabp_jcas})}.}
\label{fig:bijcas_schematic}
\end{figure}
\end{minipage}
\hfill
\begin{minipage}[H]{0.65\textwidth}
\vspace{-1ex}
\begin{figure}[H]
\centering
\includegraphics[width=0.95\textwidth]{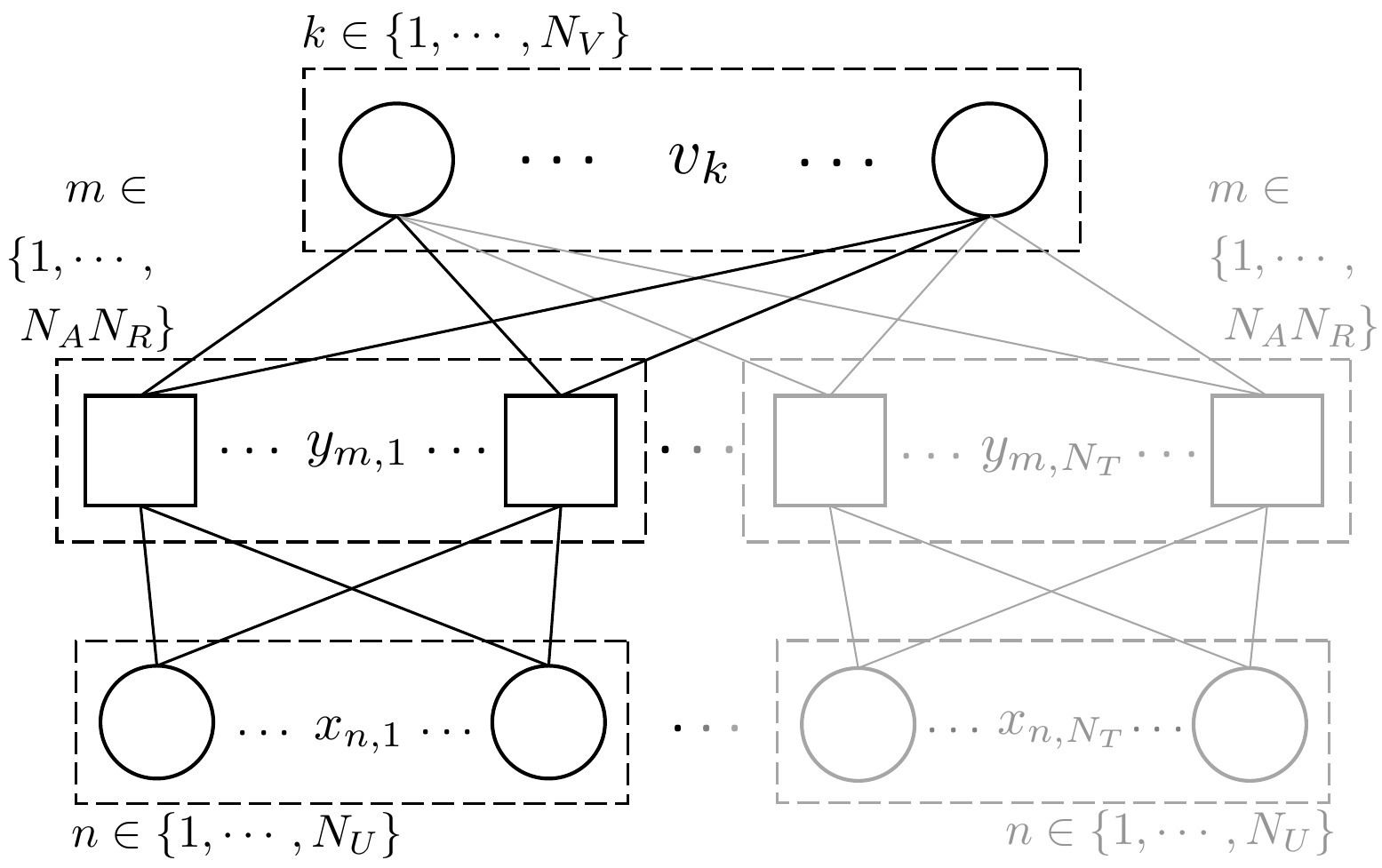}
\vspace{-1.5ex}
\caption{The tripartite factor graph of the bilinear system.}
\label{fig:factorgraph_bilinear}
\end{figure}
\end{minipage}

\vspace{3ex}
\noindent nodes) are the received symbols, and the two sets of variable nodes (circular nodes) corresponds to the environment vector $\mathbf{v}$ and the signal matrix $\mathbf{X}$, respectively.
An important distinction is made between the two types of variable nodes, which is that a data variable node receives messages only from $N_A N_R$ factor nodes corresponding to the same time instance $t$, while an environment variable node receives messages from all $N_A N_R N_T$ factor nodes.

We highlight the higher complexity of the factor graph in Fig. \ref{fig:factorgraph_bilinear} compared to those of linear \ac{GaBP} schemes shown in Figs. \ref{fig:factorgraph_linear_v} and \ref{fig:factorgraph_linear_x}, due to the asymmetric and embedded system structure of equation \eqref{eq:full_system} in relation to both variables together, as opposed to those of linear systems where one variable is considered at a time.
Other than that, the messages transferred over the graph edges are the same information as in the linear \ac{GaBP} in Section \ref{sec:ALISAC}, \textit{i.e.,} the soft-replicas, \acp{MSE}, and conditional \acp{PDF} of the variables of interest.
However, since neither of the latter is known, except as soft-replicas, the corresponding calculation of the messages must incorporate the uncertainties in \ul{both} variables, in the form of the respective \acp{MSE}.

In light of the above, similarly to Section \ref{sec:ALISAC}, the exchanged messages are constructed on the basis of soft-replicas of the variables.
Since the entries in $\mathbf{X}$ corresponding to $t \in \{1, \cdots, N_P\}$ are pilot symbols $\mathbf{X}_P$, the corresponding soft-replicas are set to their known values, \textit{i.e.,} $\hat{x}_{n,t:m,t} = (x_{p})_{n,t} ~ \forall t \in \{1, \cdots, N_P\}$, with the corresponding \ac{MSE} values set to $0$.
The remaining soft-replicas and \acp{MSE} for $t \in \{N_P + 1, \cdots, N_T\}$ are as given in equations \eqref{eq:linear_gabp_v_mse} and \eqref{eq:linear_gabp_x_mse}.

{\color{black}The complete description of the proposed \ac{Bi-ISAC} algorithm is summarized in the form of a pseudocode in Algorithm \ref{alg:bi_gabp_jcas}, and the corresponding equations of the message passing rules are elaborated in the following.}

\begin{algorithm}[H]
\hrulefill
\begin{algorithmic}[1] 
\vspace{-1.5ex}
\setlength{\baselineskip}{18pt}%
\Statex \hspace{-3ex} {\bf{Inputs:}} Received signal matrix $\mathbf{Y}$, channel matrices $\mathbf{H}$, $\mathbf{A}$, and $\mathbf{B}$, pilot matrix $\mathbf{X}_P$, noise variance $N_0$, prior distribution of environment and transmit symbols $\mathbb{P}_{\mathsf{v}_{k}}(v_{k})$ and $\mathbb{P}_{\mathsf{x}_{n,t}}(x_{n,t})$.  \vspace{-3.5ex}
\Statex \hspace{-3ex} {\bf{Outputs:}} Estimated environment vector $\tilde{\mathbf{v}}$ and estimated data signal matrix {\color{black}$\tilde{\mathbf{X}}_D$}. \vspace{-2.5ex}
\Statex \hspace{-4ex}\hrulefill \vspace{-1ex}
\Statex \hspace{-3.5ex} \textit{For data variable nodes corresponding to the pilot block, i.e., for} $t \in \{1, \cdots, N_P\}$,
\State Initialize the soft-replicas as pilots by $\hat{x}_{n,t:m,t} = (x_p)_{n,t};$ 
\Comment{$\forall m,n$}
\State Initialize the \acp{MSE} $\psi^x_{n,t:m,t}$ to $0$;
\Comment{$\forall m,n$}
\Statex \hspace{-3.5ex} \textit{For data variable nodes corresponding to the data block, i.e., for} $t \in \{N_P + 1, \cdots, N_T\}$,
\State Initialize the soft-replicas as $\hat{x}_{n,t:m,t} = \Exp_{\mathsf{x}_{n,t}}[x_{n,t}];$ 
\Comment{$\forall m,n$}
\State Initialize the \acp{MSE} $\psi^x_{n,t:m,t}$ via \eqref{eq:linear_gabp_x_mse}; \Comment{$\forall m,n$}
\Statex \hspace{-3.5ex} \textit{For all environment variable nodes, i.e., for} $t \in \{1, \cdots, N_T\}$,
\State Initialize the environment soft-replicas as $\hat{v}_{k:m,t} = \Exp_{\mathsf{v}_k}[v_k]$;  \Comment{$\forall m,k$}
\State Initialize the \acp{MSE} $\psi^v_{k:m,t}$ via \eqref{eq:linear_gabp_v_mse}; \Comment{$\forall m,k$}
\Statex \hspace{-1ex} \textbf{Until} termination criteria is satisfied$^{\color{black}*}$ \textbf{do} 
\State \hspace{1ex} Compute the soft-\ac{IC} signals $\bar{y}_{k:m,t}^{v}$ and $\bar{y}_{n,t:m,t}^{x}$ via  \eqref{eq:bigabp_softic_v_and_x}; \Comment{$\forall m,n,k,t$}
\State \hspace{1ex} Compute the conditional variances ${\nu}^{v}_{k:m,t}$ and ${\nu}^{x}_{n,t:m,t}$ via \eqref{eq:bigabp_condvar_v_and_x}; \Comment{$\forall m,n,k,t$}
\State \hspace{1ex} Compute the extrinsic mean ${\mu}^{v}_{k:m,t}$ and variance ${\varPsi}^{v}_{k:m,t}$ via \eqref{eq:bi_v_exmean_and_var}; \Comment{$\forall m,n,k,t$}
\State \hspace{1ex} Compute the extrinsic mean ${\mu}^{x}_{n,t:m,t}$ and variance ${\varPsi}^{x}_{n,t:m,t}$  via \eqref{eq:bi_x_exmean_and_var}; \Comment{$\forall m,n,k,t$}
\State \hspace{1ex} Compute the new soft-replicas $\hat{v}_{k:m,t}$ and $\hat{x}_{n,t:m,t}$ via \eqref{eq:linear_gabp_v_updatedsr} and \eqref{eq:linear_gabp_x_updatedsr}; \Comment{$\forall m,n,k,t$}
\State \hspace{1ex} Compute the new \acp{MSE} $\psi^v_{k:m,t}$ and $\psi^x_{n,t:m,t}$ via \eqref{eq:linear_gabp_v_updatedmse} and \eqref{eq:linear_gabp_x_updatedmse}; \Comment{$\forall m,n,k,t$}
\State \hspace{1ex} Update all soft-replicas and \acp{MSE} via damping \cite{Som_ITW10}; \Comment{$\forall m,n,k,t$}
\Statex \hspace{-2ex} \textbf{end}
\State Compute the consensus \acp{PDF} with $\tilde{\mu}^v_{k}$ and $\tilde{\varPsi}^v_{k}$ via \eqref{eq:linear_gabp_v_conmean_and_var}, and $\tilde{\mu}^x_{n,t}$ and $\tilde{\varPsi}^x_{n,t}$  via \eqref{eq:linear_gabp_x_conmean_and_var}; \Comment{$\forall n,k,t$}
\State Compute the final soft-estimates $\tilde{v}_{k}$ and $\tilde{x}_{n,t}$ via \eqref{eq:linear_gabp_v_final_estimate} and \eqref{eq:linear_gabp_x_final_estimate}; \Comment{$\forall n,k,t$}
\State Output $\tilde{v}_{k}$ as final estimate; \Comment{$\forall k$}
\State Project $\tilde{x}_{n,t}$ to the symbol constellation $\mathcal{X}$; \Comment{$\forall n,~ \forall t \in \{N_P+1, \cdots, N_T\}$}
\State Output projected $\tilde{x}_{n,t}$ as final estimate; \Comment{$\forall n,~ \forall t \in \{N_P+1, \cdots, N_T\}$}
\caption[]{: Bilinear \ac{GaBP} Estimator for Environment Vector $\mathbf{v}$ and Data Matrix $\mathbf{X}_D$}
\label{alg:bi_gabp_jcas}
\end{algorithmic}
\end{algorithm}
\vspace{-5ex}
{\footnotesize \noindent \color{black}$^{*}\!$ The termination criteria can be set in accordance to Section IV-C and as described in Algorithm \ref{alg:linear_gabp_v}.}

\vspace{1.5ex}

In hand of the soft-replicas and their \acp{MSE}, the factor nodes perform soft-\ac{IC} for each variable $v_k$ and $x_{n,t}$ by following

$~$
\vspace{-7ex}

\begin{subequations}
\label{eq:bigabp_softic_v_and_x}
\begin{eqnarray}
\label{eq:bigabp_softic_v}
\bar{y}_{k:m,t} = y_{m,t} -  \sum_{u}^{N_U} \Big(h_{m,u} + \sum_{i \neq k}^{N_V} a_{m,i}  \hat{v}_{i:m,t}   b_{i,u} \Big) \hat{x}_{u,t:m,t}&& \\[-1ex]
&& \hspace{-49ex}= {v_k \Big(\! a_{m,k}\! \sum_{u}^{N_U} b_{k,u}  x_{u,t} \!\Big)} \!+\!\! \sum_{u}^{N_U} h_{m,u} (x_{u,t}\! - \!\hat{x}_{u,t:m,t})\! +\! \!\sum_{u}^{N_U}\! \sum_{i \neq k}^{N_V}\! a_{m,i} b_{i,u} \left(v_i  x_{u,t}\! -\! \hat{v}_{i:m,t}   \hat{x}_{u,t:m,t} \right) \!+\! w_{m,t},
\nonumber
\end{eqnarray}
\vspace{-3ex}
\begin{eqnarray}
\label{eq:bigabp_softic_x}
\bar{y}_{n,t:m,t} = y_{m,t} -  \sum_{u \neq n}^{N_U} \Big(h_{m,u} + \sum_{i}^{N_V} a_{m,i}  \hat{v}_{i:m,t}  b_{i,u} \Big) \hat{x}_{u,t:m,t}&&\\[-1ex]
&& \hspace{-52ex}=\! x_{n,t} \Big(\!h_{m,n} \!+\!\! \sum_{i}^{N_V} a_{m,i} v_{i} b_{i,n}\! \Big) \!\!+\!\!\! \sum_{u \neq n}^{N_U} h_{m,u} (x_{u,t}\! -\! \hat{x}_{u,t:m,t})\!+\!\! \sum_{u \neq n}^{N_U}\! \sum_{i}^{N_V}  \!a_{m,i} b_{i,u} \!\left(v_i  x_{u,t}\! -\! \hat{v}_{i:m,t}   \hat{x}_{u,t:m,t} \right)\! +\! w_{m,t},
\nonumber
\end{eqnarray}
\end{subequations}

\noindent where the soft-\ac{IC} for the data variables given in \eqref{eq:bigabp_softic_x} is only performed for unknown variable nodes with indices $t \in \{N_P + 1, \cdots, N_T\}$.

Following the soft-\ac{IC}, the respective conditional \acp{PDF} are become
\begin{equation}
\mathbb{P}_{\bar{\mathsf{y}}^{\mathsf{v}}_{k:m,t}}(\bar{y}^{v}_{k:m,t}|v_{k}) \propto \mathrm{exp} \Big[ \!-\!\frac{|\bar{y}^{v}_{k:m,t} -  \big( a_{m,k} \sum_{u=1}^{N_U} b_{k,u} \! \cdot \! \hat{x}_{u,t:m,t} \big) v_k |^2}{\nu^{v}_{k:m,t}} \;\Big],
\label{eq:bigabp_condpdf_v}
\end{equation}
\vspace{-2ex}
\begin{equation}
\mathbb{P}_{\bar{\mathsf{y}}^{\mathsf{x}}_{n,t:m,t}}(\bar{y}^{x}_{n,t:m,t}|x_{n,t}) \propto \mathrm{exp} \Big[ \!-\!\frac{|\bar{y}^{x}_{n,t:m,t} - \big(h_{m,n} + \sum_{i=1}^{N_V} a_{m,i} \! \cdot \! \hat{v}_{i:m,t}  \! \cdot \! b_{i,n} \big) x_{n,t}  |^2}{\nu^{x}_{n,t:m,t}} \;\Big],
\label{eq:bigabp_condpdf_x}
\end{equation}
with the respective conditional variances $\nu^{v}_{k:m,t}$ and $\nu^{x}_{n,t:m,t}$ given by
\vspace{-1.5ex}
\begin{subequations}
\label{eq:bigabp_condvar_v_and_x}
\begin{align}
\label{eq:bigabp_condvar_v}
{\nu^{v}_{k:m,t}} &= \Exp_{\{\mathsf{v_{k}}, \mathsf{x_{n,t}}\}} \Big[ |\bar{y}^{v}_{k:m,t} -  \big( a_{m,k} \sum_{u}^{N_U} b_{k,u} \! \cdot \! \hat{x}_{u,t:m,t} \big) v_k |^2 \Big]\\[-2ex]
&= E_{v} \!\cdot\! |a_{m,k}|^2 \sum_{u}^{N_U} |b_{k,u}|^2 \psi^x_{u,t:m,t} + \sum_{u}^{N_U} |h_{m,u}|^2 \psi^x_{u,t:m,t} \nonumber \\[-1ex]
& ~~~ + \sum_{u}^{N_U} \sum_{i \neq k}^{N_V} |a_{m,i}|^2 \Big( \psi^v_{i:m,t}|\hat{x}_{u,t:m,t}|^2 + \psi^x_{u,t:m,t}(|\hat{v}_{i:m,t}|^2 + \psi^v_{i:m,t}) \Big) |b_{i,u}|^2 + N_0,\nonumber 
\end{align}
\vspace{-4ex}
\begin{align}
\label{eq:bigabp_condvar_x}
{\nu^{x}_{n,t:m,t}} &= \Exp_{\{ \mathsf{v_{k}}, \mathsf{x_{n,t}}\}} \Big[ |\bar{y}^{x}_{n,t:m,t} - \big(h_{m,n} + \sum_{i}^{N_V} a_{m,i} \! \cdot \! \hat{v}_{i:m,t}  \! \cdot \! b_{i,n} \big) x_{n,t}  |^2 \Big] \\[-2ex]
& = E_{\mathcal{X}} \cdot \! \sum_{i}^{N_V} (|a_{m,i}|^2 \psi^v_{i:m,t} |b_{i,n}|^2) + \sum_{u \neq n}^{N_U}(|h_{m,u}|^2 \psi^x_{u,t:m,t}) \nonumber \\[-1ex]
& ~~~ + \sum_{u \neq n}^{N_U} \sum_{i}^{N_V} |a_{m,i}|^2 \Big( \psi^v_{i:m,t}|\hat{x}_{u,t:m,t}|^2 + \psi^x_{u,t:m,t}(|\hat{v}_{i:m,t}|^2 + \psi^v_{i:m,t}) \Big) |b_{i,u}|^2 + N_0,\nonumber 
\end{align}
\end{subequations}

\vspace{-0.5ex}
\noindent where the expectation $E_{v} \triangleq \Exp_{\{ \mathsf{v_{k}}\}}[|v_{k}|^2]$ has been introduced for convenience of notation.

In turn, all variable nodes compute the interference-cancelled extrinsic belief \acp{PDF} given by
\vspace{-0.5ex}
\begin{equation}
\mathbb{P}_{\mathsf{l}^{\mathsf{v}}_{k:m,t}}(\ell^{v}_{k:m,t} | v_k) = \! \! \prod_{p \neq m}^{N_A N_R} \!\prod_{q \neq t}^{N_T} \; \mathbb{P}(\bar{y}^{v}_{k:p,q}|v_{k}) \propto \mathrm{exp} \left[ -\frac{|v_k - \mu^v_{k:m,t}|^2}{\varPsi^v_{k:m,t}} \right],
\label{eq:bi_v_expdf}
\end{equation}
\vspace{-1ex}
\begin{equation}
\mathbb{P}_{\mathsf{l}^{\mathsf{x}}_{n,t:m,t}}(\ell^{x}_{n,t:m,t} | x_{n,t}) = \prod_{p \neq m}^{N_A N_R} \mathbb{P}(\bar{y}^{x}_{n,t:p,t}|x_{n,t}) \propto \mathrm{exp} \left[ -\frac{|x_n - \mu^x_{n,t:m,t}|^2}{\varPsi^x_{n,t:m,t}} \right],
\label{eq:bi_x_expdf}
\vspace{-0.5ex}
\end{equation}

\noindent with the respective extrinsic means and variances given by 
\vspace{1ex}
\begin{subequations}
\label{eq:bi_v_exmean_and_var}
\begin{equation}
\mu^v_{k:m,t} = \varPsi^v_{k:m,t} \cdot \Big(\sum\nolimits_{p \neq m}^{N_A N_R}  \sum\nolimits_{q \neq t}^{N_T}\!\frac{ (a_{p,k} \sum_{u=1}^{N_U} b_{k,u} \! \cdot \! \hat{x}_{u,q:p,q})^* \cdot \bar{y}^{v}_{k:p,q}}{ \nu^{v}_{k:p,q}}\Big),
\label{eq:bi_v_exmean}
\end{equation}
\vspace{-2ex}
\begin{equation}
\varPsi^v_{k:m,t} = \Big(\sum\nolimits_{p \neq m}^{N_A N_R}  \sum\nolimits_{q \neq t}^{N_T} \!\frac{|a_{p,k} \sum_{u=1}^{N_U} b_{k,u} \! \cdot \! \hat{x}_{u,q:p,q}|^2}{  \nu^{v}_{k:p,q}} \Big)^{\!\!-1}\!\!\!\!,
\label{eq:bi_v_exvar}
\end{equation}
\end{subequations}
\vspace{-2ex}
\begin{subequations}
\label{eq:bi_x_exmean_and_var}
\begin{equation}
\mu^x_{n,t:m,t} = \varPsi^x_{n,t:m,t} \cdot \Big(\sum\nolimits_{p \neq m}^{N_A N_R} \frac{ (h_{p,n} + \mathsmaller\sum_{i=1}^{N_V} a_{p,i} \! \cdot \! \hat{v}_{i:p,t}  \! \cdot \! b_{i,n})^* \cdot \bar{y}^{x}_{n,t:p,t}}{ \nu^{x}_{n,t:p,t}}\Big),
\label{eq:bi_x_exmean}
\end{equation}
\vspace{-2ex}
\begin{equation}
\varPsi^x_{n,t:m,t} = \Big(\sum\nolimits_{p \neq m}^{N_A N_R} \frac{|h_{p,n} + \sum_{i=1}^{N_V} a_{p,i} \! \cdot \! \hat{v}_{i:p,t}  \! \cdot \! b_{i,n}|^2}{  \nu^{x}_{n,t:p,t}} \Big)^{\!\!-1}\!\!\!\!.
\label{eq:bi_x_exvar}
\end{equation}
\end{subequations}
\vspace{-4ex}

Using equations \eqref{eq:bi_v_exmean} through \eqref{eq:bi_x_exvar}, the updated soft-replicas and the \acp{MSE} can be obtained following the same operations as in the linear \ac{GaBP}, namely, equations \eqref{eq:linear_gabp_v_updatedsr} and \eqref{eq:linear_gabp_v_updatedmse} for $v_k$, and equations \eqref{eq:linear_gabp_x_updatedsr} and \eqref{eq:linear_gabp_x_updatedmse} for $x_{n,t}$, with the post-convergence consensus as in equations \eqref{eq:linear_gabp_v_conmean_and_var} and \eqref{eq:linear_gabp_v_final_estimate} for $\hat{v}_{k:m,t}$, and equations \eqref{eq:linear_gabp_x_conmean_and_var} and \eqref{eq:linear_gabp_x_final_estimate} for $\hat{x}_{n,t:m,t}$, respectively.

\vspace{-1ex}
\section{Performance Evaluation}
\label{sec:performance}

\vspace{-1.5ex}
\subsection{Complexity Analysis}
\label{sec:complexity}
\vspace{-1ex}

In Table \ref{tab:complexities}, the complexity orders of the two proposed algorithms and their constituent modules are given, in terms of the system size parameters $N_U, N_V, N_A, N_R, N_P, N_T$, with $\rho \triangleq N_P/N_T$ denoting the pilot length ratio, and $\lambda$ denoting the number of algorithm iterations {\color{black} which is assumed to be equal between the two proposed algorithms, for comparison.}

\begin{table}[b]
\vspace{-6ex}
\centering
\caption{\color{black}Complexity orders the proposed and the \ac{SotA} algorithms.}
\label{tab:complexities}
\vspace{-2ex}
\small
\begin{tabular}{|l||c|}
\hline
Linear \ac{GaBP} on $\mathbf{v}$ (Algorithm \ref{alg:linear_gabp_v}), only pilot phase & $\mathcal{O} \big[\lambda N_U (N_V N_A N_R N_P)^2 \big]$ \\
\hline
Linear \ac{GaBP} on $\mathbf{X}$ (Algorithm \ref{alg:linear_gabp_x}), only data phase & $\mathcal{O} \big[\lambda (N_U N_A N_R N_D)^2 \big]$ \\
\hline
Linear \ac{GaBP} on $\mathbf{v}$ (Algorithm \ref{alg:linear_gabp_v}), pilot and data phase & $\mathcal{O} \big[\lambda N_U (N_V N_A N_R N_T)^2 \big]$ \\
\hline
\multirow{2}{*}{Full \ac{AL-ISAC} algorithm (Algorithm \ref{alg:alt_gabp_jcas})} & $\mathcal{O} \big[\lambda (N_U N_V N_A N_R)^2 \big( \frac{N_{\!P}^{\;2} + N_T^{\;2}}{N_U} + N_{\!D}^{\;2} \big) \big]$ \\
&  $= \mathcal{O} \big[\lambda (N_U N_V N_A N_R N_T)^2 \big( (1 - \rho)^2 + \frac{\rho^2 + 1}{N_U}\big)  \big]$ \\
\hline
Full \ac{Bi-ISAC} algorithm (Algorithm \ref{alg:bi_gabp_jcas}) & $\mathcal{O} \big[\lambda(N_U N_V N_A N_R N_T)^2]$ \\
\hline
\hline
\color{black} \Acf{SotA} SCMA-IRS-MPA scheme\footnotemark \cite{Tong_JSTSP21} & \color{black} $\mathcal{O}\big[ R d_f M^{d_f} + N_U M + \lambda^2 N_U N_V N_A N_R \big]$ \\ 
\hline
\end{tabular}
\vspace{-4ex}
\end{table}  

\footnotetext{\color{black} Unlike the proposed methods, the reference \ac{SotA} scheme of \cite{Tong_JSTSP21}  is tied to a \ac{SCMA} design \cite{Taherzadeh_VTC14}, as can be inferred from system parameters such as $R, d_f, M$, such that a direct comparison in terms of complexity is not possible.
We emphasize, however, that the proposed \ac{AL-ISAC} and \ac{Bi-ISAC} are shown to outperform the \ac{SotA} in both the environment estimation \ac{MSE} and \ac{BER} under the same SNR, as discussed in Section and shown Fig. \ref{fig:sota_comp}. \vspace{-2.5ex}}

\begin{figure}[t]
\centering
\includegraphics[width=0.54\textwidth]{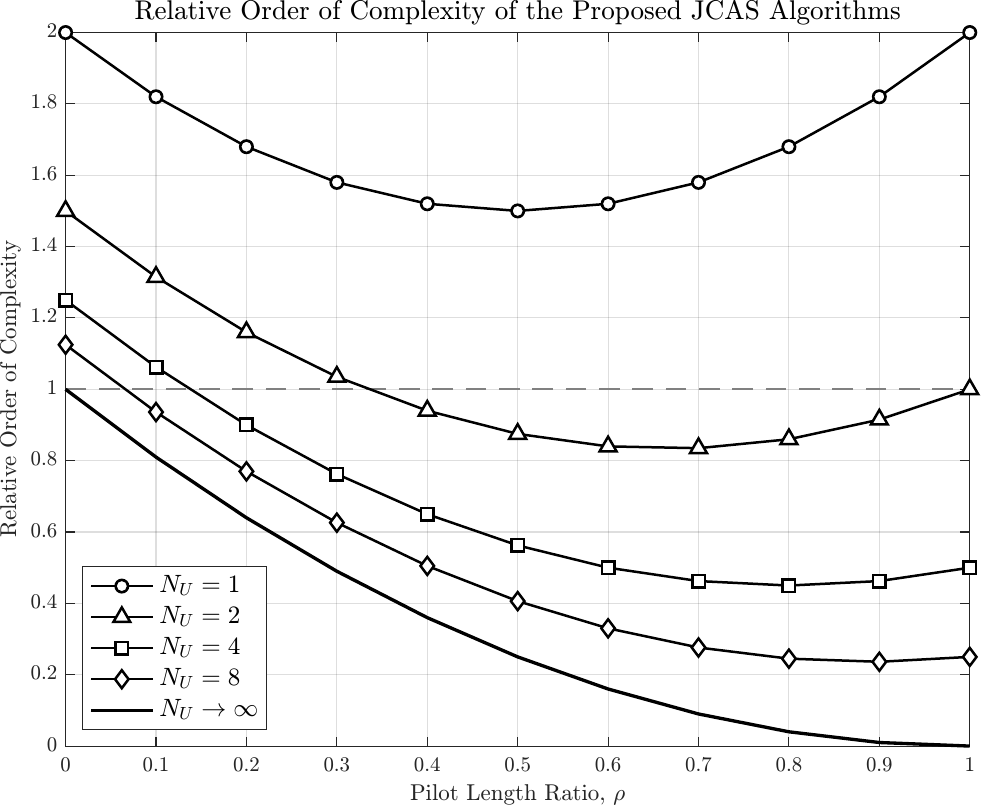}
\vspace{-1ex}
\caption{The relative order of complexity (given by $\frac{\text{complexity of AL-ISAC}}{\text{complexity of Bi-ISAC}}$) of the two proposed algorithms for varying values of $\rho$ and $N_U$.}
\label{fig:complexity_comp}
\vspace{-4.5ex}
\end{figure}

The comparison of the full order of complexity shows that the \ac{Bi-ISAC} has a second order complexity dependent on all system size parameters and a linear complexity on the number of iterations.
Interestingly, it is found that the \ac{AL-ISAC} requires the same order of complexity as the \ac{Bi-ISAC}, except for the scaling factor $(1 - \rho)^2 + \frac{\rho^2 + 1}{N_U}$, which is dependent on $N_U$ and $\rho$.
The latter is therefore a measure of the relative complexity of the two proposed algorithms, which is plotted in Fig. \ref{fig:complexity_comp}, for various values of $N_U$ as a function of $\rho$. 

Considering first the case of $N_U = 1$, it is seen that the relative complexity is larger than 1 for all values of pilot length ratio $\rho \in [0,1]$, indicating that in a single-\ac{UE} case, the \ac{AL-ISAC} algorithm requires a higher complexity over the \ac{Bi-ISAC} algorithm, regardless of amount of pilot symbols.
It is also seen, however, that in the multi-\ac{UE} scenario ($N_U \geq 2$), both algorithms have the same complexity order if pilot length ratio $\rho_{eq} = \frac{N_{U} - |\sqrt{N_{U} \;^{\!\!2} - N_{U} - 1}|}{N_U + 1}$; and further, that when $\rho > \rho_{eq}$, the relative complexity decreases below 1, indicating that the \ac{AL-ISAC} achieves a lower complexity than the \ac{Bi-ISAC} with increasing $N_U$ and sufficient $\rho$.

Fig. \ref{fig:complexity_comp} also indicates that the relative complexity follows a truncated quadratic behavior with the minimum value occuring at the specific pilot ratio $\rho_{min} = \frac{N_U}{N_U + 1}$, which converges to 1 for $N \rightarrow \infty$.
The result implies that a sufficiently high pilot length ratio is required for the \ac{AL-ISAC} algorithm to achieve the optimal complexity, especially with increasing $N_U$.
In conclusion, the \ac{Bi-ISAC} algorithm enjoys a complexity that is robust to the number of \acp{UE} and the length of pilots (throughput), whereas the \ac{AL-ISAC} algorithm has a complexity that decreases with the number of \acp{UE}, but at the cost of throughput.

\vspace{-2ex}
\subsection{Performance Comparison Against the State-of-the-Art}
\label{sec:proposed_vs_sota}
\vspace{-0.5ex}

In this section, we compare the performance of the two proposed algorithms against the \ac{SotA} method{\color{black}\footnote{\color{black}To the best of our knowledge, the method in \cite{Tong_JSTSP21} is the only \ac{SotA} \ac{ISAC} method based on a similar voxelated model.}} from \cite{Tong_JSTSP21}, which also utilizes a voxelated grid to perform \ac{ISAC} via leveraging multiple linear \ac{MP} algorithms.
We highlight, however, that the latter reference utilizes the support of a fully known \ac{IRS} in the \ac{ROI}, and strongly relies on sparsity in the received signal, offered by means of an \ac{SCMA} interface between the \acp{UE} and the \ac{AP}, while our contributions are not limited to such a conditions.
Instead, the proposed methods operate over fully dense received signals and with no support of \acp{IRS}.

Due to this distinction in system set-up, an adequate system parametrization must be considered for a fair comparison of our proposed algorithms against the \ac{SotA} method of \cite{Tong_JSTSP21}.
Specifically, in the \ac{SCMA} scheme of \cite{Tong_JSTSP21}, each single-antenna \ac{UE} transmits an $M$-bit code by utilizing $d_f$ out of $R$ orthogonal frequency bands, which is received by a single \ac{AP} with $N_R$ receive antennas.
Since the considered system for our proposed algorithms is a single-frequency model \eqref{eq:received_signal}, the diversity gain between each \ac{UE} and the \ac{CPU} is mimicked by setting the number of \acp{AP} as $N_A = d_f$, such that  $N_A N_R = d_f \cdot N_R$, such that the number of nodes and edges of the final factor graph is the same in both systems.

\begin{figure}[b]
\vspace{-5ex}
\centering
\begin{subfigure}[t]{0.49\textwidth}
\centering
\includegraphics[width=\textwidth]{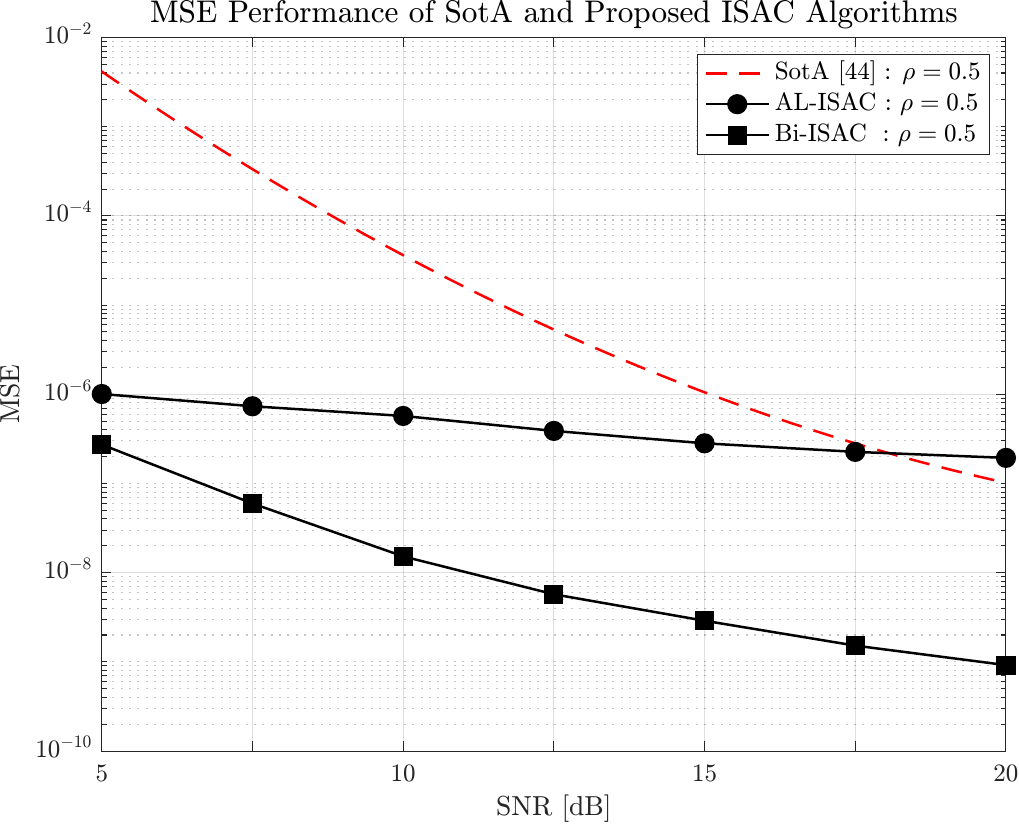}
\vspace{-5ex}
\caption{\Ac{MSE} performance.}
\label{fig:sota_comp_mse}
\end{subfigure}
\hfill
\begin{subfigure}[t]{0.49\textwidth}
\centering
\includegraphics[width=\textwidth]{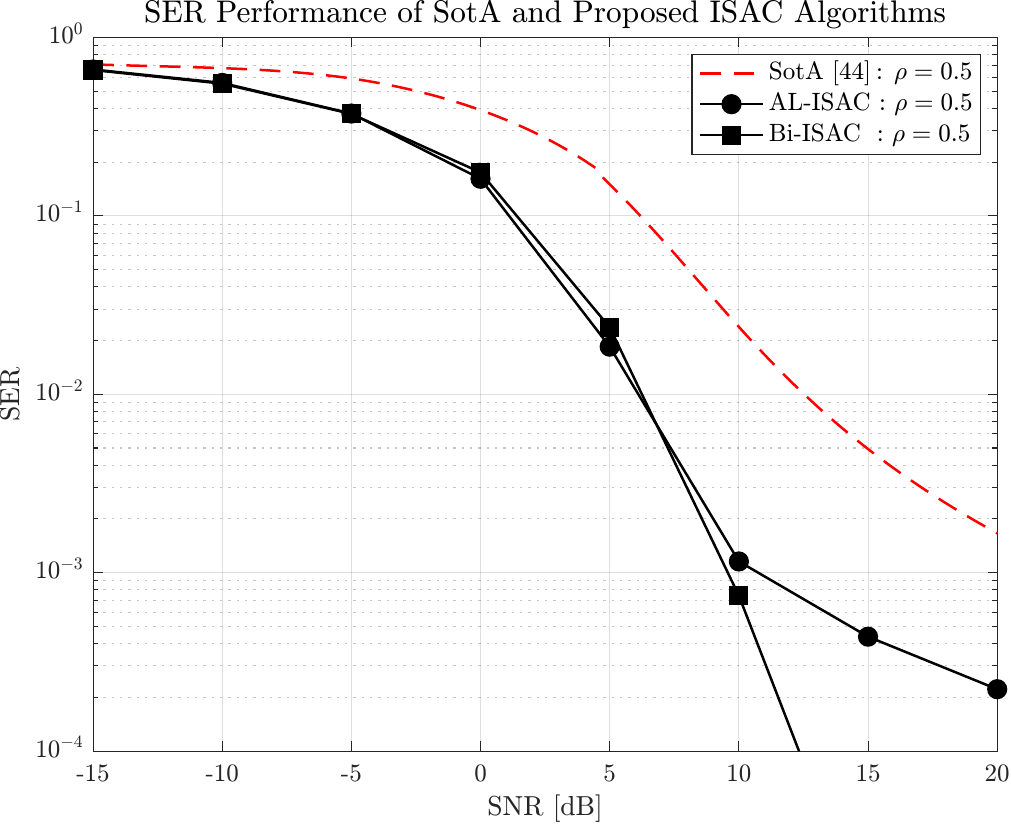}
\vspace{-5ex}
\caption{\Ac{SER} performance.}
\label{fig:sota_comp_ser}
\end{subfigure}
\vspace{-2ex}
\caption{\ac{MSE} and \ac{SER} performances of the \ac{SotA} \cite{Tong_JSTSP21} against the proposed algorithms, with $N_U = 6$, $N_R = 9$, $N_A = d_f = 2$, $N_V = 8 \times 8 \times 8$, $E_v = 1.5\%$, $M = 4$, $N_T = 100$, and $\rho = 0.5$.}
\label{fig:sota_comp}
\vspace{-3ex}
\end{figure}

Fig. \ref{fig:sota_comp} illustrates the sensing and communication performances of the two proposed \ac{ISAC} algorithms in comparison to the \ac{SotA} \ac{ISAC} algorithm of \cite{Tong_JSTSP21}, in terms of the \ac{MSE} and \acp{SER}, respectively.
First, in Fig. \ref{fig:sota_comp_mse}, the sensing performance, $i.e.$, the \ac{MSE} of the voxel occupancy coefficients, is evaluated.
Under equivalent system parameters, in particular with a pilot ratio\footnote{The value $\rho = 0.5$ is taken from \cite{Tong_JSTSP21} in order to enable direct comparison.} of $\rho = 0.5$, the \ac{MSE} of the proposed \ac{Bi-ISAC} algorithm is found to significantly outperform that of the \ac{SotA} method at all \ac{SNR} values, while the \ac{AL-ISAC} is found to be slightly outperformed by the latter at the high \acp{SNR} regime.

Then, in Fig. \ref{fig:sota_comp_ser}, the communication performances of the \ac{ISAC} systems are evaluated in terms of the \ac{SER} of the estimated symbols.
It can be seen that both proposed algorithms exhibit superior symbol estimation performances compared to the \ac{SotA}, with the \ac{Bi-ISAC} exhibiting the additionally desirable feature that no error floors are observed even at higher \acp{SNR}.

All in all, the results corroborate the claim that both proposed methods generally outperform the \ac{SotA} method of \cite{Tong_JSTSP21} in both sensing and communication functionalities.

\vspace{-2ex}
{\color{black}
\subsection{Convergence Behavior of the Proposed Algorithms}
\label{sec:convergence_analysis}
\vspace{-0.5ex}

In view of the superior performance of the two proposed \ac{ISAC} algorithms in comparison with the \ac{SotA} \cite{Tong_JSTSP21}, we proceed in this section to further analyze the two proposed \ac{ISAC} algorithms in detail, aiming also to clarify advantages of each relative to the other.
To that extent, we first study the convergence behavior of the proposed \ac{ISAC} algorithms so as to obtain insight on the appropriate \ac{MP} termination criteria and damping parameters to be utilized.

\begin{figure}[b]
\vspace{-3ex}
\centering
\begin{subfigure}[t]{0.49\textwidth}
\centering
\includegraphics[width=\textwidth]{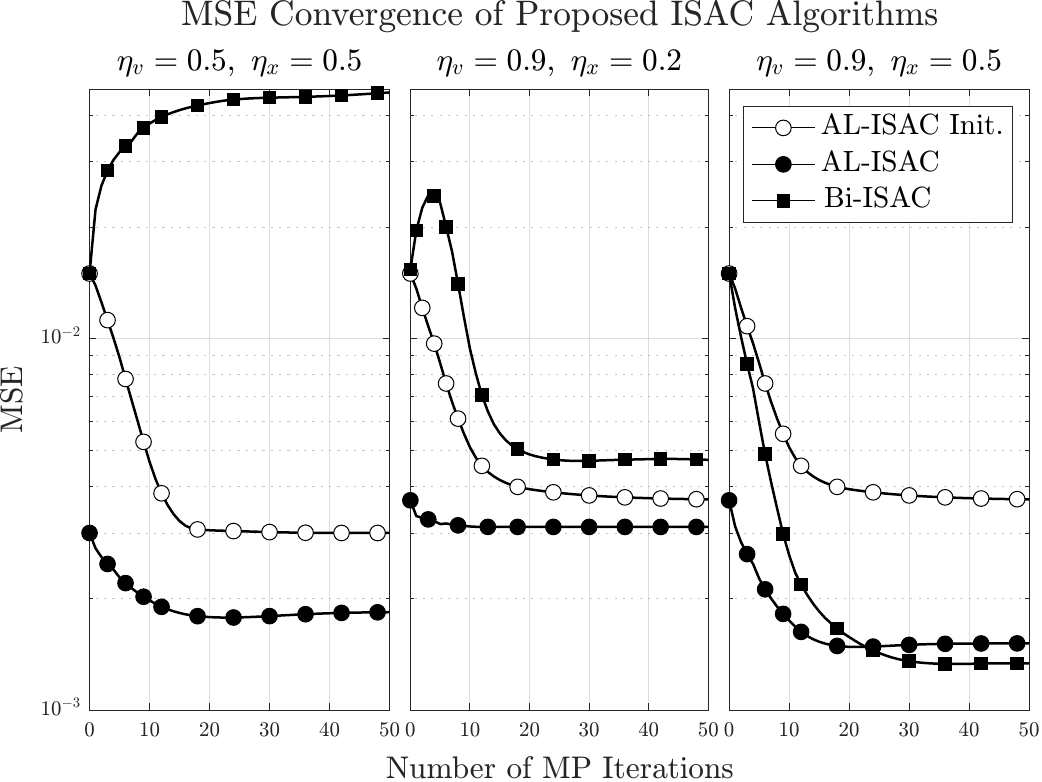}
\vspace{-4.5ex}
\caption{\color{black}\ac{MSE} convergence.}
\label{fig:mse_conv}
\end{subfigure}
\hfill
\begin{subfigure}[t]{0.49\textwidth}
\centering
\includegraphics[width=\textwidth]{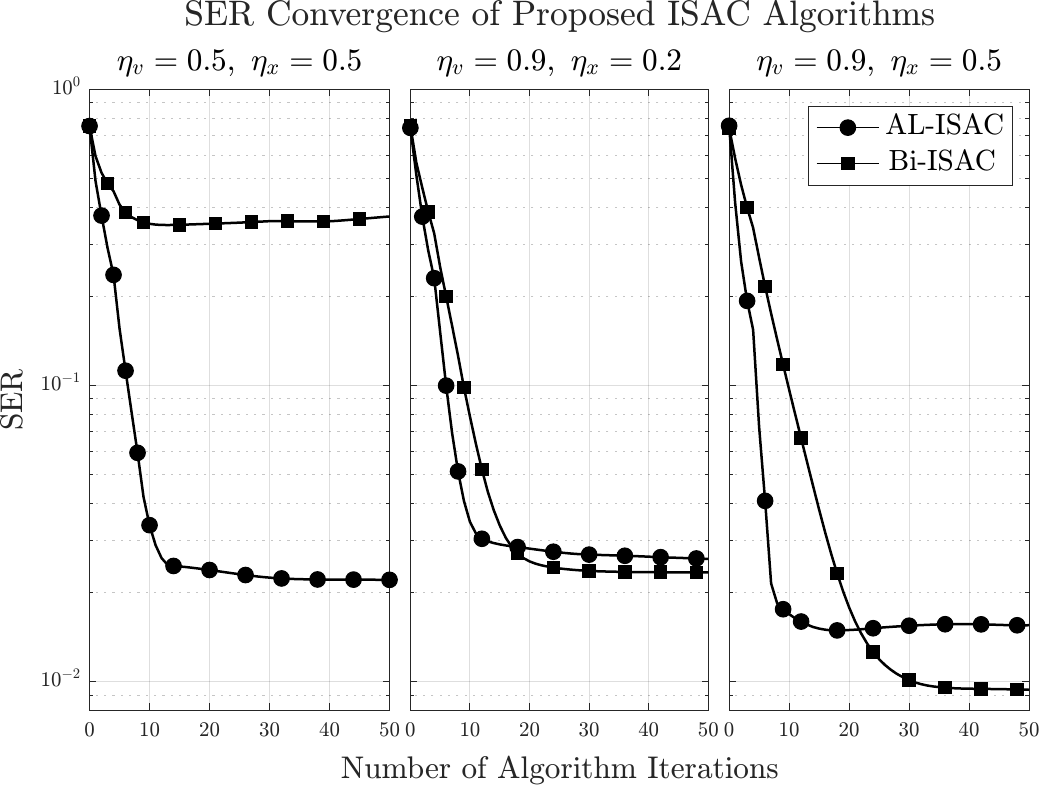}
\vspace{-4.5ex}
\caption{\color{black}\ac{SER} convergence.}
\label{fig:ser_conv}
\end{subfigure}
\vspace{-2ex}
\caption{\color{black}Convergence behavior of the proposed \ac{AL-ISAC} and \ac{Bi-ISAC} algorithms with varying damping factors $\eta_x$ and $\eta_v$ with $\rho\!=\!0.1, ~E_v\!=\!1.5\%$, $\text{SNR}\!=\! 15\text{dB}$, as a function of \ac{MP} iterations in a system with $N_U = 4, ~N_AN_R = 12, ~N_V = 512$, and $N_T = 100$.}
\label{fig:conv_perf}
\end{figure}

Figure \ref{fig:conv_perf} depicts the convergence behavior of the proposed \ac{ISAC} algorithms in terms of the \ac{MSE} of the voxel coefficient soft-replicas $\hat{v}_{k:m,t}$ and the \ac{SER} of the data symbol soft-replicas $\hat{x}_{n,t:m,t}$, for three selected combinations of their respective damping parameters $\eta_x \in [0,1]$ and $\eta_v \in [0,1]$, following the damped update rule given by \cite{Som_ITW10} \vspace{-0.3ex}
\begin{equation}
\hat{x}_{n,t:m,t}^{(\tau)} \overset{\text{update}}{\longleftarrow} \eta_x \!\cdot\! \hat{x}_{n,t:m,t}^{(\tau -1)} + (1 - \eta_x) \!\cdot\! \hat{x}_{n,t:m,t}^{(\tau)}, ~~ \text{and}~~ \hat{v}_{k:m,t}^{(\tau)} \overset{\text{update}}{\longleftarrow} \eta_v \!\cdot\! \hat{v}_{k:m,t}^{(\tau -1)} + (1-\eta_v) \!\cdot\! \hat{v}_{k:m,t}^{(\tau)}, \vspace{-0.3ex}
\end{equation}
where the superscript $[\;\!\cdot\;\!]^{(\tau)}$ denotes the soft-replica at the $\tau$-th iteration of the \ac{MP} loop.

It can be observed in Fig. \ref{fig:mse_conv} that while different values of $\eta_v$ all result in convergence of \ac{MSE}, the convergent value is sensitive to the parameterization of $\eta_v$, especially for the \ac{Bi-ISAC} which suffers from high \ac{MSE} with low damping, as opposed to the \ac{AL-ISAC} which is less prone to such errors in both the initial (white circles) and the final (black circles) estimation.
Similar and consequent results in the \ac{SER} are observed in Fig. \ref{fig:ser_conv}, where convergence is also achieved in all cases, but the convergent value is largely affected by the \ac{MSE} performance and $\eta_v$, while the effect of $\eta_x$ is not as prominent to the final result compared to $\eta_v$.

In light of the above results, the parameterization $\eta_x = 0.5$ and $\eta_v = 0.9$ and the convergence criteria of $\lambda = 100$  iterations is selected in the following simulations to ensure a reliable convergence behavior{\color{black}\footnote{\color{black}In principle, the optimal damping parameters resulting in the best \ac{MSE} and \ac{SER} performances can be heuristically searched for each simulation set up, but this is not considered here due to the prohibitive complexity.}}.}

\vspace{-2ex}
\subsection{Robustness Analysis of the Proposed Algorithms}
\label{sec:robustness_performance_analysis}
\vspace{-0.5ex}
To that end, we shall utilize performance metrics for the sensing and communication functions of \ac{ISAC} systems other than the \ac{MSE} of the estimated voxel coefficients and the \ac{SER} of the estimated communication symbols, which were used in \cite{Tong_JSTSP21, Tao_WCSP20, Zhang_ICC2022} and the previous section.

For the sensing function in particular, we remark that metrics used for radar-based \ac{ISAC} cannot be used directly due to the unique voxelated occupancy grid-based approach followed here.
It is therefore sensible to instead introduce a new metric, referred to as the \ac{VOER}, which measures the rate of \ac{FP} and \ac{FN} elements, defined as the incorrect estimation of an occupied voxel element in presence of an empty ground-truth, and the incorrect estimation of an empty voxel element in the presence of an occupied ground-truth, respectively.
Mathematically, the \ac{VOER} is therefore defined as $\Exp\big[||\mathbf{v} - \tilde{\mathbf{v}}||_{0}\big]/N_V$, where $\mathbf{v}$ is the ground truth, $\tilde{\mathbf{v}}$ is the estimate vector, while $\|\cdot\|_{0}$ denotes the $\ell_0$-norm of a vector.

Notice that for the trivial all-empty (or ``blind'') estimator, which returns $\tilde{\mathbf{v}} = \mathbf{0}_{N_V \times 1}$, the \ac{VOER} reduces to $E_v \triangleq \Exp[||\mathbf{v}||_0]/N_V = \mathrm{VOER}_{empty}$, which is the average sparsity of the environment.
We can therefore utilize this figure as an absolute reference of performance, in the sense $\mathrm{VOER} \ll \mathrm{VOER}_{empty}$ indicates a good sensing performance by a given \ac{ISAC} method.

Finally, instead of the \ac{SER} often used in related literature, we opt to evaluate the communication performance of the proposed \ac{ISAC} schemes in terms of the more descriptive \ac{BER}, defined as $\mathrm{BER} \triangleq \Exp[B_e]/B$, where $B_e$ denotes the number of errorneously detected data bits of $\mathbf{X}_D$, and $B$ is the total number of bits conveyed in $\mathbf{X}_D$.

{\color{black}

\begin{figure}[H]
\vspace{-1ex}
\centering
\begin{subfigure}[t]{0.497\textwidth}
\centering
\includegraphics[width=\textwidth]{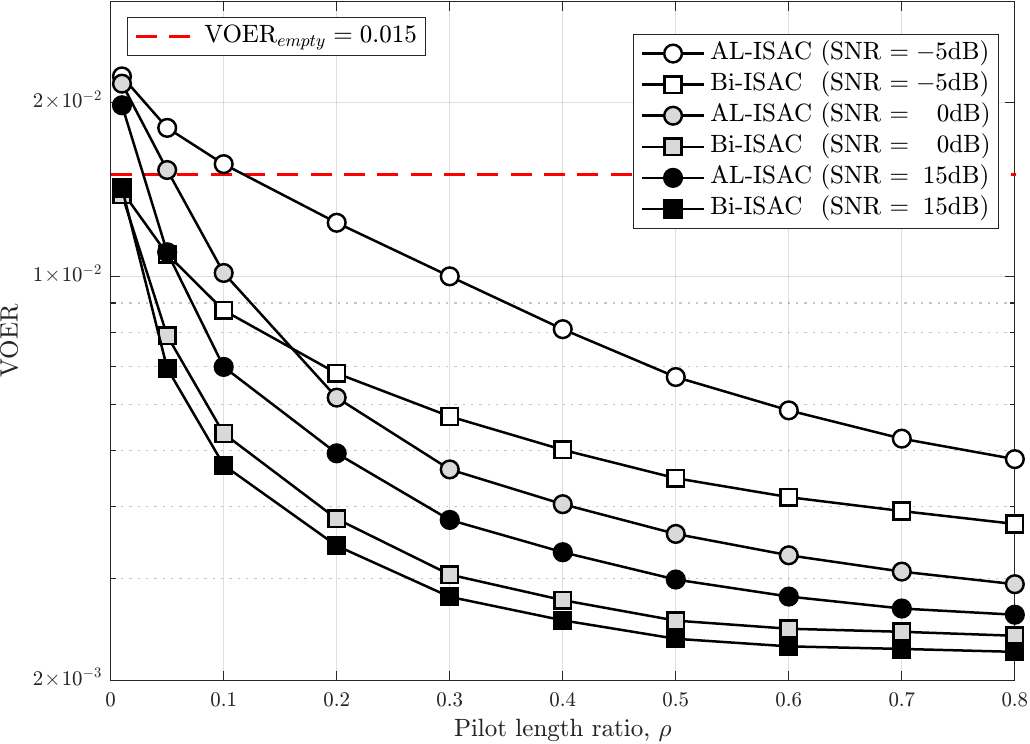}
\vspace{-4.5ex}
\caption{\color{black}Effect of pilot ratio $\rho$ for different \acp{SNR}, with fixed environment sparsity $E_v = 1.5\%$ and \ac{SNR} $= 15\mathrm{dB}$.}
\label{fig:voer_with_rho}
\end{subfigure}
\hfill
\begin{subfigure}[t]{0.483\textwidth}
\centering
\includegraphics[width=\textwidth]{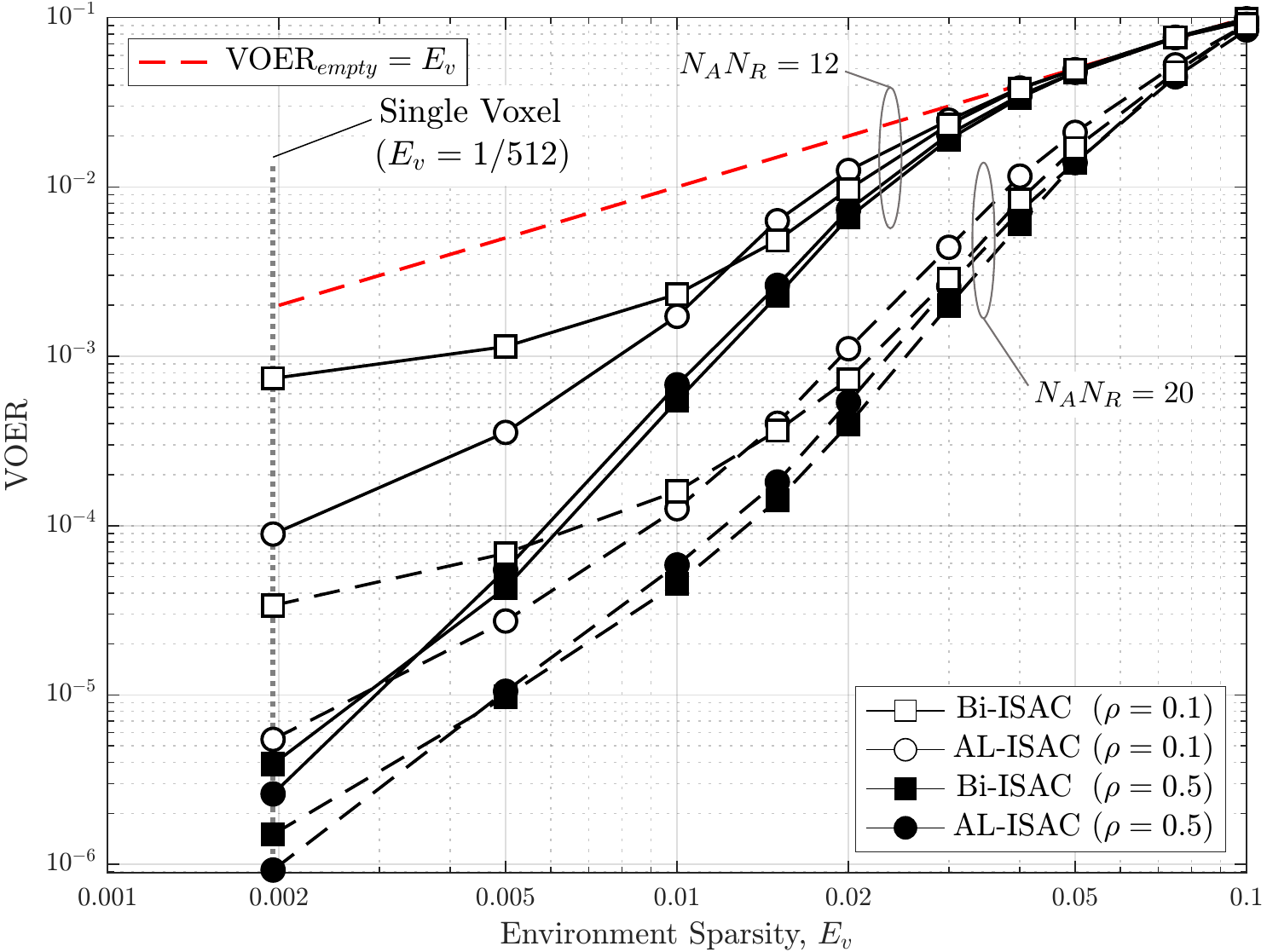}
\vspace{-4.5ex}
\caption{\color{black}Effect of environment sparsity $E_v$ with varying pilot ratio $\rho$, with fixed \ac{SNR} $=15\mathrm{dB}$.}
\label{fig:voer_with_ev}
\end{subfigure}
\vspace{-1.5ex}
\caption{\color{black} \ac{VOER} performance of the proposed \ac{AL-ISAC} and \ac{Bi-ISAC} algorithms in a system with $N_U = 4, ~N_AN_R = 12, ~N_V = 512$, and $N_T = 100$.}
\label{fig:VOER_perf} 
\vspace{-3.5ex}
\end{figure}

First, we evaluate the robustness of the two proposed algorithms in terms of their \ac{VOER} performances, as a function of different system parameters including the pilot ratio $\rho$, \ac{SNR}, and environment sparsity $E_v$ (voxel occupancy probability). 

In Fig. \ref{fig:voer_with_rho}, the effect of varying the pilot ratio $\rho$ is illustrated for various \acp{SNR}. The results show} that, for a specific value of $\rho$, the \ac{Bi-ISAC} outperforms the \ac{AL-ISAC} for the entire \ac{SNR} range; or alternatively, that for a given \ac{SNR}, the \ac{Bi-ISAC} scheme requires a much lower number of pilots to achieve the same \ac{VOER} performance of the \ac{AL-ISAC} method.
{\color{black}
Next, Fig. \ref{fig:voer_with_ev} depicts the effect of environment sparsity $E_v$ on the \ac{VOER} performance of the proposed algorithms, where it is observed that the \ac{AL-ISAC} can actually achieve a superior performance over the \ac{Bi-ISAC} in extremely sparse environments as can be seen in the single voxel case of $E_v = 1/512 \approx 0.002$.
On the other hand, the \ac{Bi-ISAC} is more robust in the sense that it exhibits smaller gradient, which suggests that the \ac{Bi-ISAC} is less affected by changes in $E_v$. The \ac{Bi-ISAC} also tends to outperform the \ac{AL-ISAC} as $E_v$ increases (\textit{i.e.,} for $E_v > 0.01$).

\begin{figure}[b]
\vspace{-4ex}
\centering
\begin{subfigure}[t]{0.49\textwidth}
\centering
\includegraphics[width=\textwidth]{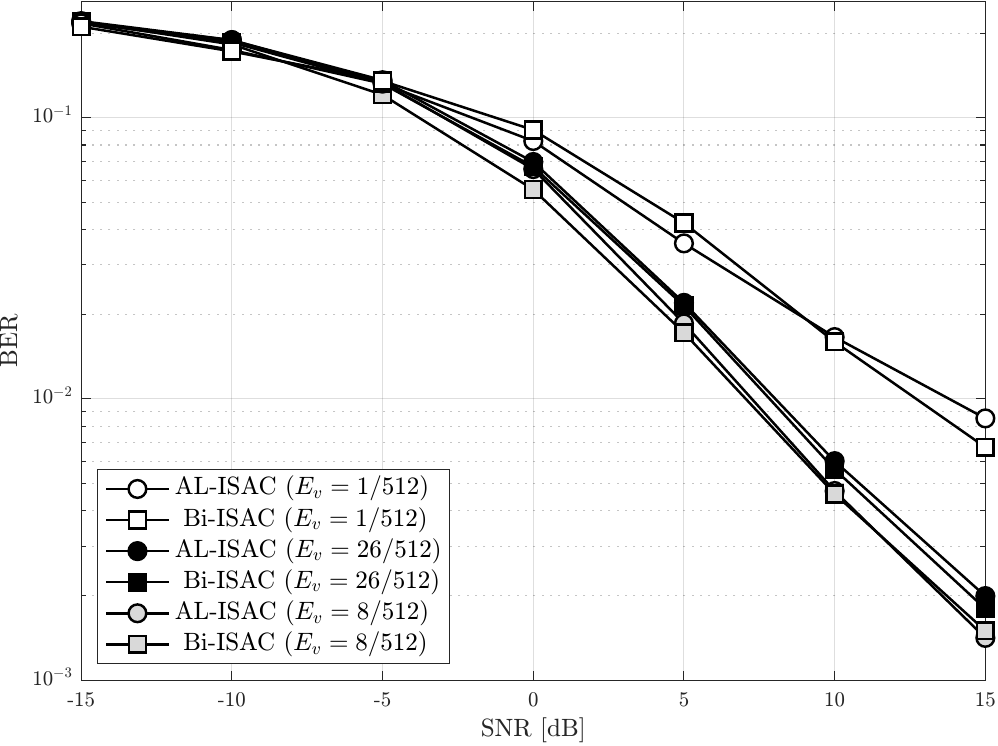}
\vspace{-4.5ex}
\caption{\color{black}Effect of environment sparsity $E_v$ and \ac{SNR}, with fixed pilot ratio $\rho = 0.1$.}
\label{fig:ber_with_snr}
\end{subfigure}
\hfill
\begin{subfigure}[t]{0.49\textwidth}
\centering
\includegraphics[width=\textwidth]{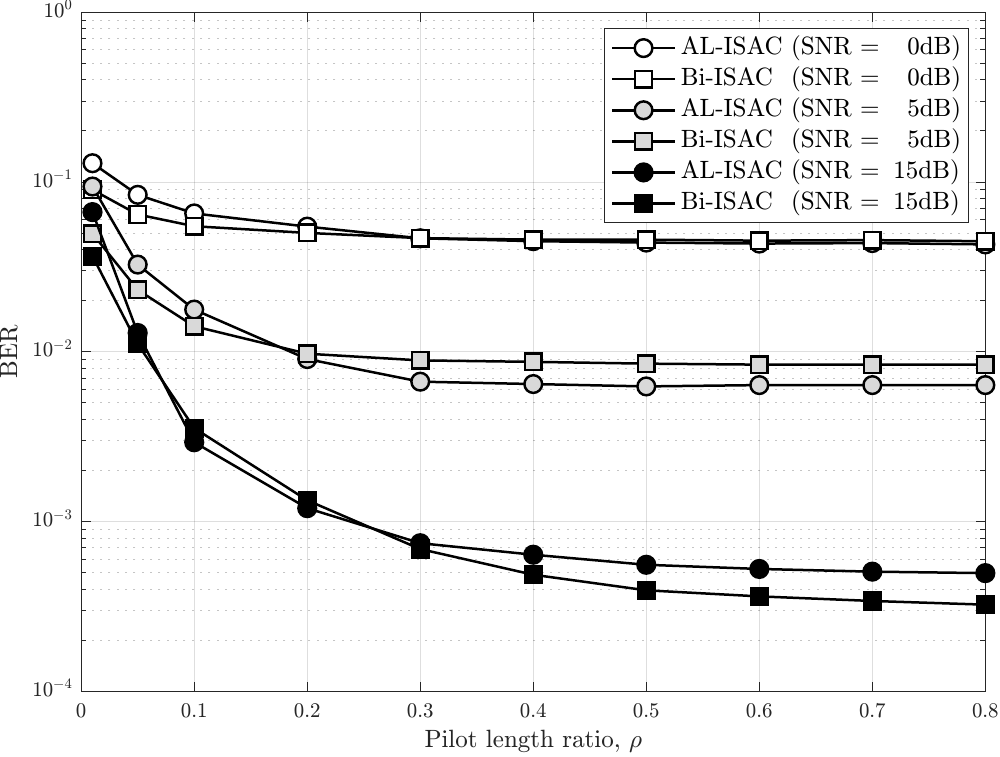}
\vspace{-4.5ex}
\caption{\color{black}Effect of pilot ratio $\rho$ for different \acp{SNR}, with fixed environment sparsity $E_v = 1.5\%$.}
\label{fig:ber_with_rho}
\end{subfigure}
\vspace{-1.5ex}
\caption{\color{black} \ac{BER} performance of the proposed \ac{AL-ISAC} and \ac{Bi-ISAC} algorithms in a system with $N_U = 4, ~N_AN_R = 12, ~N_V = 512$, and $N_T = 100$.}
\label{fig:BER_perf}
\vspace{-3ex}
\end{figure}

Next, we proceed to evaluate the \ac{BER}  performance of the proposed algorithms, which are compared in Fig. \ref{fig:ber_with_snr} as a function of the \ac{SNR} and for different environment sparsities, namely $E_v = 0.002, 0.015, 0.05$, which respectively correspond to $1, 8,$ and $26$ occupied voxels (out of $512$ total) in the environment.
Since the \ac{VOER} performance as seen in Fig. \ref{fig:VOER_perf} degrades with larger $E_v$, one may expect the \ac{BER} performance to follow the same behavior.
However, it is observed that an extremely sparse environment results in \ac{BER} performance degradation.

While counter-intuitive at a first sight, these results are actually intuitive if analyzed from an information/estimation-theoretical viewpoint.
Although fundamental limits on the performances of ISAC are still to be derived, it is to be expected that with a certain amount of resources such as power, side information ($i.e.$, pilot signals) and computational complexity, a fundamental limit on the joint estimation and communications performances exists.
Indeed, from an algorithmic viewpoint, a large number of voxels implies that much of the degrees-of-freedom are utilized for environment estimation, which negatively impacts the communications performance.
Similarly, an extremely sparse environment implies that most voxel coefficients are $0$, such that the corresponding connected edges in the factor graph are nullified, deteriorating \ac{BER} performance.}

{\color{black} Next, Fig. \ref{fig:ber_with_rho} evaluates the effect of the pilot ratio on the \ac{BER} performance, where} it can be seen that for small pilot ratios (\textit{i.e.,} $\rho < 0.2$), the two proposed algorithms achieve a similar performance, but that for larger pilot ratios (\textit{i.e.,} $\rho > 0.3$), the \ac{AL-ISAC} outperforms the \ac{Bi-ISAC} in moderate SNR cases (\textit{i.e.,} $5\mathrm{dB}$).
This result, which can be counter-intuitive to the reader, is actually expected and explained by the fact that linear \ac{MP} modules of the \ac{AL-ISAC} scheme are constructed on the assumption of perfect symbol knowledge (0 uncertainty for symbol estimates), whose assumptions are increasing met with large pilot ratios.
Ultimately, however, at significantly high \acp{SNR} (\textit{i.e.,} SNR $\geq 15\mathrm{dB}$), the \ac{Bi-ISAC} is again shown to outperform \ac{AL-ISAC}, which is a direct consequence of the error-floor behavior exhibited by the \ac{AL-ISAC} algorithm.

\begin{figure}[H]
\vspace{-1ex}
\centering
\begin{subfigure}[t]{0.497\textwidth}
\centering
\includegraphics[width=\textwidth]{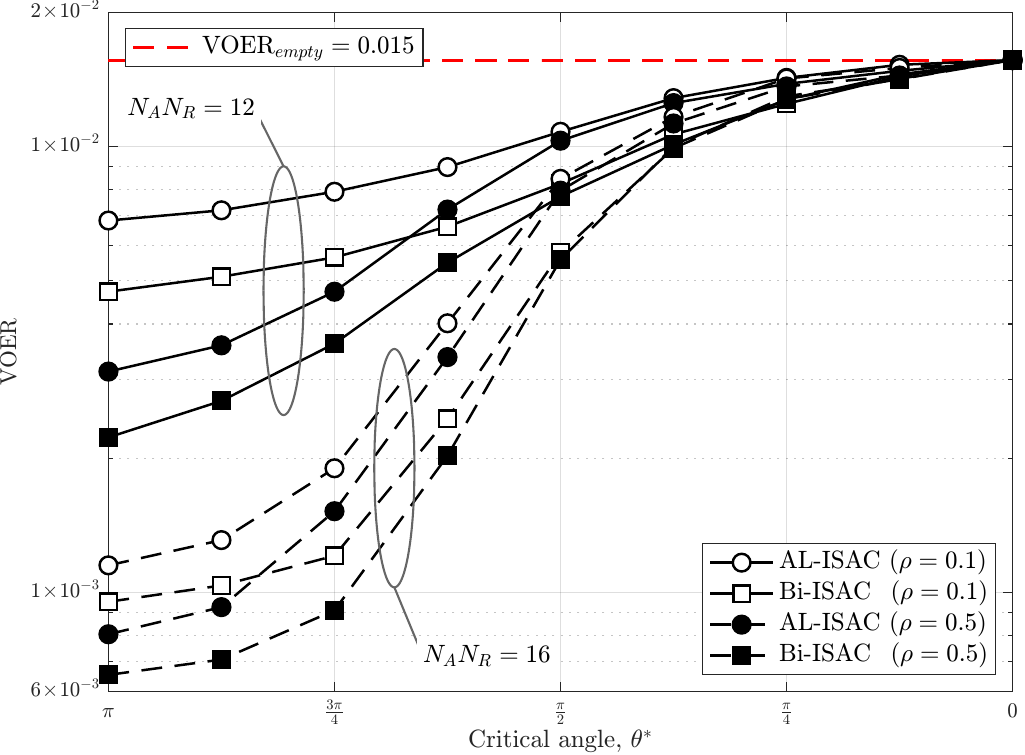}
\vspace{-5ex}
\caption{\ac{VOER} performance.}
\label{fig:voer_with_punc}
\end{subfigure}
\hfill
\begin{subfigure}[t]{0.483\textwidth}
\centering
\includegraphics[width=\textwidth]{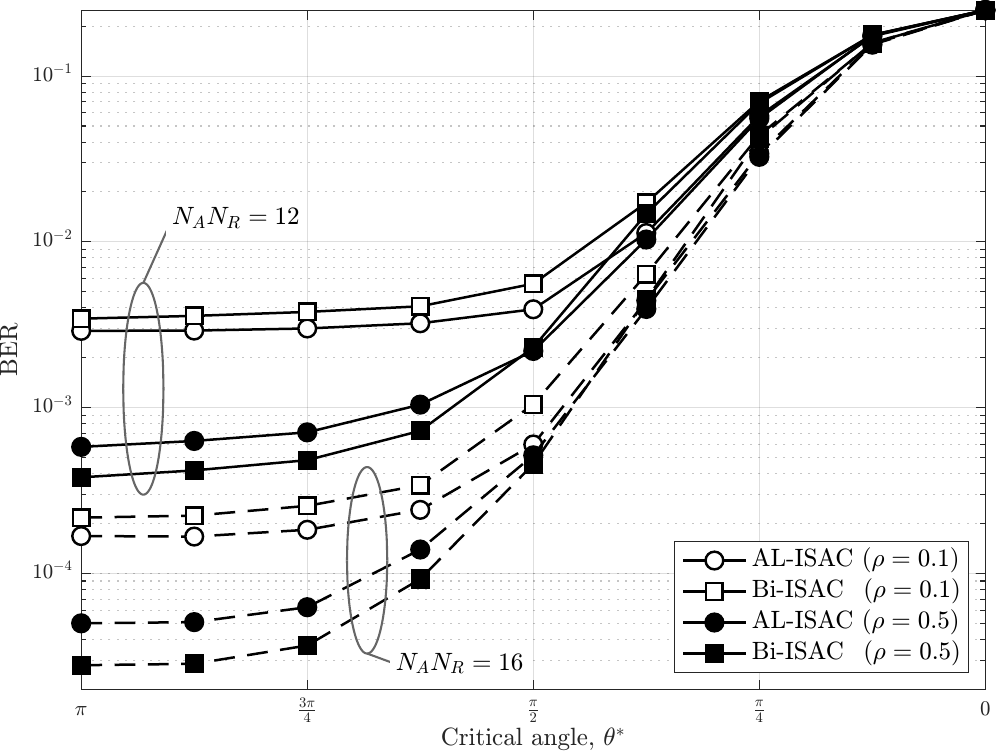}
\vspace{-5ex}
\caption{\ac{BER} performance.}
\label{fig:ber_with_punc}
\end{subfigure}
\vspace{-2ex}
\caption{\color{black}Performance of the proposed \ac{AL-ISAC} and \ac{Bi-ISAC} algorithms with varying pilot ratios $\rho$, with $E_v = 1.5\%,~ \text{SNR} = 15\text{dB}$, as a function of the critical channel blockage angle $\theta^*$.}
\label{fig:perf_with_punc}
\vspace{-3ex}
\end{figure}

Finally, Fig. \ref{fig:perf_with_punc} elucidates the effect of random channel blockages as discussed in Section \ref{sec:environment_model}. 
In particular, the figure compares the \ac{VOER} and \ac{BER} performances of the two proposed \ac{ISAC} algorithms with respect to the critical angle $\theta^*$, which determines the channel blockage rate following the stochastic-geometric empirical model derived in Sec. \ref{sec:environment_model} (see Fig. \ref{fig:angle_dist_PDF}).

The environment sensing performance illustrated in Fig. \ref{fig:voer_with_punc} exhibits a similar behavior to the effect of $\rho$, where the \ac{Bi-ISAC} achieves a superior performance for all cases.
In addition, the \ac{Bi-ISAC} curves exhibit a slower increase in gradient compared to those of \ac{AL-ISAC}, which indicates the higher robustness of \ac{Bi-ISAC} to path blockages.
The superior robustness of the \ac{Bi-ISAC} algorithm against both pilot length and random channel blockages can be accredited to the increased number of edges in the full factor graph arising from the bilinear representation of the system, which can be seen by comparing Fig. \ref{fig:factorgraph_linear_v} and Fig. \ref{fig:factorgraph_linear_x} of the linear case against Fig. \ref{fig:factorgraph_bilinear} of the bilinear case.
Each factor node of the bilinear factor graph is connected to a significantly larger number of variable nodes as compared to the linear factor graphs, which implies more remaining edges for stable message passing even when a large number of edges are removed.
\ac{MP} over the pruned graph is only feasible when there are sufficient pilot data still connected to the main graph, therefore making the \ac{Bi-ISAC} more dependent on the pilot ratio for stability.

As for the communications performances, compared in Fig. \ref{fig:ber_with_punc}, it is found that the behavior of both schemes differ with the pilot ratio.
For low pilot ratios, the \ac{AL-ISAC} is shown to slightly outperform the \ac{Bi-ISAC}, whereas for high pilot ratios, the \ac{Bi-ISAC} exhibits a more significantly superior performance to the \ac{AL-ISAC}, further corroborating the results of Fig. \ref{fig:voer_with_rho}.

$~$
\vspace{-7ex}
\section{Conclusion}

We proposed two new \ac{ISAC} schemes in which a voxelated \ac{3D} representation of a \ac{ROI} is extracted from the scattering features present in the effective \ac{CSI}, utilizing the same physical layer communications air interface of an uplink connection between multiple single-antenna \acp{UE} and multi-antenna \acp{AP}.
The first scheme, dubbed \ac{AL-ISAC}, relies on a modular feedback structure in which the transmit data and the environment are estimated alternately, whereas the second method, referred to as \ac{Bi-ISAC}, leverages the bilinear inference framework to estimate both variables concurrently.
Both contributed methods were shown via computer simulations to outperform the \ac{SotA} in accurately recovering the transmitted data, as well as in obtaining a voxelated \ac{3D} image of the environment. 
An analysis of the computational complexities and robustness of the proposed methods revealed distinct advantages of each scheme, namely, that \ac{Bi-ISAC} exhibits an overall best performance and robustness to short pilots and channel blockages, while \ac{AL-ISAC} offers lower complexity especially in scenarios with large numbers of \acp{UE}{\color{black}, and can exhibit superior performance under ideal conditions such as long pilot blocks, high environment sparsity, and no channel blockages.}



\vspace{-1.5ex}

\end{document}